# Audio-visual cross-modality knowledge transfer for machine learning-based in-situ monitoring in laser additive manufacturing

Jiarui Xie [a], Mutahar Safdar [a], Lequn Chen [b,c,*], Seung Ki Moon [b], Yaoyao Fiona Zhao [a,*]

[a] Department of Mechanical Engineering, McGill University, Montreal, Quebec, Canada, H3A 0G4
[b] School of Mechanical and Aerospace Engineering, Nanyang Technological University, Singapore, 639798
[c] Advanced Remanufacturing and Technology Center (ARTC), Agency of Science, Technology and Research (A*STAR) Singapore, 637143
[*] Corresponding authors.
Email addresses: yaoyao.zhao@mcgill.ca (Y. F. Zhao) and chen1470@e.ntu.edu.sg (L. Chen)

## Abstract

Various machine learning (ML)-based in-situ monitoring systems have been developed to detect anomalies and defects in laser additive manufacturing (LAM) processes. While multimodal fusion, which integrates data from visual, audio, and other modalities, can improve monitoring performance, it also increases hardware, computational, and operational costs. This paper introduces a cross-modality knowledge transfer (CMKT) methodology for LAM in-situ monitoring, which transfers knowledge from a source modality to a target modality. CMKT enhances the representativeness of the features extracted from the target modality, allowing the removal of source modality sensors during prediction. This paper proposes three CMKT methods: semantic alignment, fully supervised mapping, and semi-supervised mapping. The semantic alignment method establishes a shared encoded space between modalities to facilitate knowledge transfer. It employs a semantic alignment loss to align the distributions of identical groups (e.g., visual and audio defective groups) and a separation loss to distinguish different groups (e.g., visual defective and audio defect-free groups). The two mapping methods transfer knowledge by deriving features from one modality to another using fully supervised and semi-supervised learning approaches. In a case study for LAM in-situ defect detection, the proposed CMKT methods were compared with multimodal audio-visual fusion. The semantic alignment method achieved an accuracy of 98.6% while removing the audio modality during the prediction phase, which is comparable to the 98.2% accuracy obtained through multimodal fusion. Using explainable artificial intelligence, we discovered that semantic alignment CMKT can extract more representative features while reducing noise by leveraging the inherent correlations between modalities.

## Nomenclature

| Abbreviation | Full phrase |
| --- | --- |
| AM | Additive manufacturing |
| LAM | Laser additive manufacturing |
| ML | Machine learning |
| KNN | K-nearest neighbors |
| LPBF | Laser powder bed fusion |

| IoT | Internet of Things |
| CNN | Convolutional neural network |
| LDED | Laser direct energy deposition |
| LoF | Lack of fusion |
| WAAM | Wire arc additive manufacturing |
| CMKT | Cross-modality knowledge transfer |
| CCD | Charge-coupled device |
| ROS | Robot operating system |
| RKHS | Reproducing kernel Hilbert space |
| TP | True positive |
| TN | True negative |
| FP | False positive |
| FN | False negative |
| TPR | True positive rate |
| FPR | False positive rate |
| TNR | True negative rate |
| AUC | Area under the curve |
| ROC | Receiver operating characteristics |
| AUC-ROC | Area under the Receiver operating characteristics curve |
| t-SNE | T-distributed stochastic neighbor embedding |
| LIME | Local interpretable model-agnostic explanations |
| XAI | Explainable artificial intelligence |
| OM | Optical microscope |
| AWGN | Additive white Gaussian noise |

## 1. Introduction

Additive Manufacturing (AM) or 3D printing-based techniques fabricate parts in a layer-upon-layer manner as opposed to conventional manufacturing techniques based on material removal [1]. AM offers several benefits such as tool elimination, material savings, design freedom, part consolidation, prototyping ease, and mass customization. These AM features unlock constraints on product design and production efficiency and have a significant economic potential [2]. AM technologies also enable specialized applications such as the printing of multifunctional or multi-material designs. Despite their merits, AM techniques also suffer from a multitude of process defects and process anomalies limiting their adoption for large-scale industrial production [3,4]. The potential of AM to offer unique solutions to new and existing challenges in design and manufacturing has inspired a multi-faceted research landscape spanning materials, hardware, and software development [5,6].

AM process development has seen numerous research efforts focused on advanced monitoring techniques [3]. These efforts are fueled by the process complexity (e.g., hardware, software, and material aspects), range of physical scale (e.g., microscopic to macroscopic), domain diversity (e.g., spatial and/or temporal), and information magnitude (e.g., frequency, volume, and variety) [7,8]. The motivation to monitor rests in the potential to capture information associated with defective or anomalous features from process phenomena (interactions, melting, re-melting, and cooling) or physical objects (melt pools, layers, and under-build parts) for downstream analytics applications [9]. Chen et al. [10] recently reviewed in-situ monitoring techniques for laser-based AM and grouped them into optical, acoustic, laser-line scanning, and multi-sensor-based categories. Each category offers unique merits suiting diverse downstream applications such as process state prediction, defect detection or classification, and property or performance estimation [9]. The rise of multi-sensor fusion techniques hints at the potential of overcoming the limitations faced by individual monitoring approaches and opens the possibility of knowledge transfer across information modalities [11].

The in-situ data collected from process monitoring systems can be used to train machine learning (ML) models for various AM tasks, including process anomaly detection, defect detection, and quality prediction [10]. For example, metal AM process anomaly detection models can be trained using vision-based melt pool monitoring data to identify the locations of potential defects [12]. Image segmentation models trained on labeled layer-wise AM images can classify defects at the pixel level [13]. Combined with process parameters and part geometry information, in-process data can also be used to infer the part quality such as geometric conformity and mechanical properties [14]. According to the defect detection and quality prediction results, process parameters can be adjusted to enhance the part quality. Although ML-based AM process monitoring systems offer abundant potential for improving AM process reliability, they have not been extensively deployed in real-life production environments. The costs of monitoring instruments and data acquisition might not be sufficiently compensated by the enhanced part quality [10]. Besides, the quality control standards of industrial production demand high performance from AM process modeling results.

Multisensor monitoring and data fusion have been recently implemented to enrich data informativeness for modeling performance improvement [15,16]. This method is founded on the assumption that a single sensor only captures part of the physical phenomena of the AM process; thus, the collected dataset is likely to have missing information about the modeling task. Multisensor fusion utilizes data from multiple sensors to complement each other and fill the information gap. Multimodal fusion is a special type of multisensor fusion that integrates datasets of different modalities, including vision-based, acoustic-based, and thermal-based data [17]. Multisensor and multimodal fusion have been proven effective in laser powder bed fusion (LPBF) defect detection [17,18] and quality prediction [16]. However, the improved predictive performance is accompanied by higher costs and larger data volumes, making AM process monitoring less affordable. A multisensor system considerably increases the deployment cost because of the expenses associated with purchasing, operating, and maintaining multiple sensors. Moreover, the enlarged data volume of multisensor fusion increases data handling and model prediction runtime. This requires a powerful computation capacity to ensure real-time defect detection and process control, further increasing the cost. In an Internet of Things (IoT) setting where data are wirelessly transmitted to the cloud, a large data volume leads to high power consumption and long transmission time.

The recent advancements in LAM in-situ monitoring research have focused primarily on developing sophisticated models that incorporate extensive data and complex ML architectures, often complicating real-world deployment. While multimodal fusion enhances predictive performance in LAM processes, the associated high operational and computational costs create barriers for practical deployment in production environments. In addition to providing useful information, data from multiple modalities can also introduce noise and task-irrelevant information, which might consume learning capacity and degrade model performance. To better leverage multimodal datasets for LAM process monitoring, this paper proposes a cross-modality knowledge transfer (CMKT) methodology aimed at reducing data volume and sensor costs while maintaining comparable performance. In multimodal fusion, data from multiple modalities are collected in both the training and prediction phases (Figure 1a). In contrast, CMKT only collects data from multiple modalities in the training phase, where knowledge is transferred from the source modality to the target modality (Figure 1b). The proposed methodology removes the source modality and utilizes the target modality as the input during the prediction phase, thus enhancing operational and computational efficiencies. Moreover, CMKT exploits the correlation between melt pool visual and audio modalities to extract features shared between modalities, hence neglecting noises that are usually modal-specific [19,20]. The key contributions of this paper are summarized below:

- This paper proposes a CMKT methodology that integrates in-situ laser direct energy deposition (LDED) visual and audio data and transfers knowledge between the two modalities.
- A semantic alignment CMKT method is proposed to semantically align the visual and audio modalities of in-situ LDED data in a shared encoded space.
- Two cross-modality mapping methods are proposed to conduct CMKT by deriving the features of the source modality from the target modality for in-situ LDED data.
- The CMKT methods are implemented to train LDED process defect detection models leading to higher predictive performance and operational efficiency.
- We reveal the denoising and salient feature extraction capabilities of semantic alignment CMKT using an explainable artificial intelligence (XAI) technique named linear interpretable model-agnostic explanations (LIME).

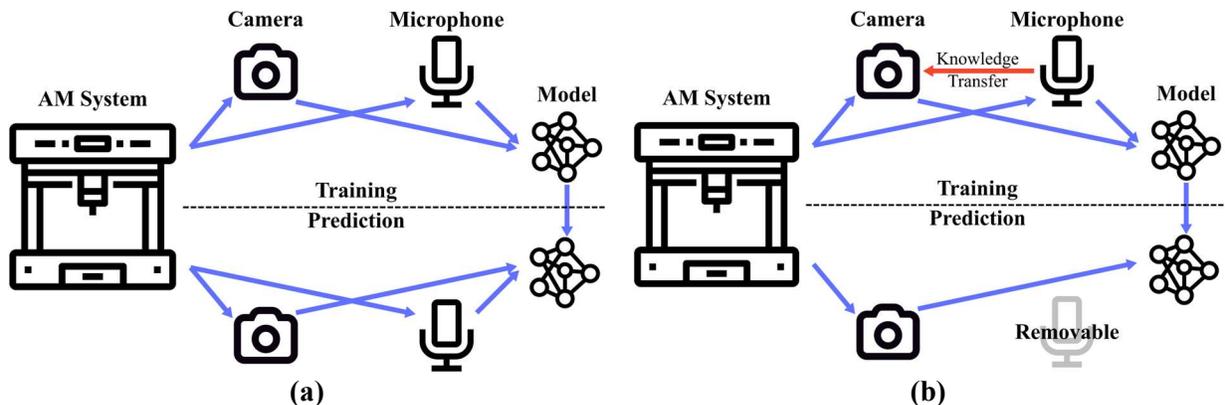

**Figure 1: System setups of (a) multimodal fusion and (b) cross-modality knowledge transfer.**

The remainder of this paper is structured as follows. Section 2 reviews AM process monitoring and knowledge transfer, highlighting key research gaps. Section 3 delves into the experiment, dataset, and proposed CMKT methodologies. Section 4 covers the training and evaluation of the CMKT and multimodal fusion models. Section 5 presents a discussion of the modeling results, focusing on predictive performance, computational complexity, and operational efficiency, with XAI implemented to demonstrate the denoising and salient feature extraction capabilities of semantic alignment CMKT. Finally, Section 6 concludes with key takeaways from this research.

## 2. Related research

This section examines the current state of ML-assisted in-situ process monitoring in LAM, emphasizing acoustic and vision-based techniques. Each sensing modality offers unique insights into the LAM process. Here, we explore their inherent limitations and introduce knowledge transfer as a potential solution to these challenges. Additionally, we extend the discussion to CMKT, illustrating its applications in other advanced fields and proposing its novel integration into LAM to improve in-situ process monitoring and defect detection.

### 2.1. ML-assisted process monitoring in LAM

Early defect detection and correction are crucial for ensuring part quality and enhancing the reliability of LAM processes. ML-assisted in-situ process monitoring plays a pivotal role in achieving this by capturing dynamic laser-material interactions within the melt pool [21–24].

Among the various monitoring technologies, vision-based monitoring is particularly notable for its ability to directly observe and analyze melt pool dynamics. The data captured through coaxial cameras can reflect essential metallurgical phenomena such as melting, cooling, and heat transfer [25–27]. Researchers have focused on extracting physics-informed features from these images as quality indicators for predicting process anomalies and defects [28–31]. Features like melt pool morphologies and temperature, influenced by input energy density, have been correlated with process stability [32]. For instance, Khanzadeh et al. [33] used a k-nearest neighbors (KNN) model to analyze melt pool geometries and predict porosity in real-time from X-ray tomography-labeled melt pool pictures with 98.44% accuracy. Similarly, Ren et al. [34] implemented a convolutional neural network (CNN) model based on high-speed synchrotron X-ray and thermal imaging to detect keyhole pores in LPBF of Ti-6Al-4V, achieving submillisecond temporal resolution and near-perfect prediction rates. Moreover, Asadi et al. [35] proposed a YOLOv8l model for in-situ melt pool segmentation of the LDED process, achieving high accuracy in predicting bead geometry. Other examples include using off-axis thermal imaging in LPBF, which demonstrates success in localized porosity prediction [36,37].

Despite these advancements, the implementation of vision-based monitoring remains expensive and time-consuming. Moreover, integrating vision sensors into existing setups often requires customized hardware designs, which increases costs and deployment complexity. In contrast, acoustic-based monitoring offers a compelling alternative. This technique utilizes acoustic signals generated by laser-material interactions, which contain rich information about processes such as melting, solidification, and defect formation [38–41]. Unlike visual sensors, acoustic monitoring does not require extensive modifications to AM equipment, facilitating easier integration and operation. Recent studies have demonstrated the potential of acoustic signals in predicting pore concentrations and classifying lack of fusion (LoF), keyhole pores, and cracks through ML [42–46].

To further enhance the robustness of process monitoring, the integration of multisensor approaches is advocated [47–52]. By combining data from different sensing modalities, multisensor models address the limitations inherent to single-sensor setups. This method leverages the strengths of each sensor type, enabling a comprehensive analysis of the complex, multi-dimensional phenomena occurring within LAM processes. For example, Li et al. [53] developed a feature-level fusion technique, in which photodiode signal and audio signal were fused using a CNN-based model to estimate layer-wise LPBF quality. Chen et al. [52] proposed a hybrid CNN model which incorporates visual and audio feature extraction streams to achieve localized quality prediction in robotic LDED. Similar research on multisensor fusion in LAM monitoring has also been conducted in [54–56]. These studies showed that combining sensing modalities enhances defect detection accuracy and process dynamics understanding.

Despite the promising results of acoustic-based monitoring, their robustness and accuracy still lag behind those of vision-based approaches, particularly in the LDED process where noise can degrade acoustic signal quality [45]. Although multisensor approaches can enhance overall prediction accuracy, such methods often require expensive sensor setups and are generally not favored by industry end-users. Recently, Liu et al. [57] proposed a method that utilizes combined acoustic and photodiode signals to infer melt pool morphological features and predict spatially-dependent LoF defects in LPBF. Similarly, visual characteristics of the melt pool in LDED can be inferred from a more cost-effective audio modality [58]. This indicates the possibility of transferring knowledge from one sensing modality to another, which could lead to even better outcomes.

## 2.2. Knowledge transfer for in-situ process monitoring in LAM

Knowledge transfer has emerged as a powerful strategy in ML, particularly useful in domains where data are scarce or expensive to collect, such as in LAM [59]. This subsection discusses various types of knowledge transfer applied for process monitoring in LAM, categorizing them based on their application scope: knowledge transfer from public datasets to domain-specific small datasets, from one material to another, and from one process to another (e.g., from LPBF to LDED).

### 2.2.1. Public to domain-specific dataset transfer

In LAM, the scarcity of large, annotated datasets often impedes the application of deep learning techniques. A pragmatic solution is to leverage knowledge transfer from large, public datasets to smaller, domain-specific ones. For example, Fischer et al. [60] utilized this approach by employing the Xception architecture, initially pretrained on ImageNet, to monitor powder bed quality in LPBF. The proposed method achieved a classification accuracy of 99.15% by distinguishing various inhomogeneities in the powder bed using high-resolution images obtained under different lighting conditions. This approach not only boosts the performance with limited data but also significantly reduces the need for extensive data collection. Similarly, Li et al. [61] demonstrated the effectiveness of knowledge transfer in enhancing in-situ quality inspection in the LPBF by using model pretrained on ImageNet and re-trained on layer-wise visual images. Zhu et al. [62] addressed the challenge of small dataset sizes in detecting surface defects in LDED, where knowledge transfer was used to pretrain a YOLOv7 model using an open-source defect dataset (NEU-DEF) and then finetuned it with a small dataset of LDED surface images. The results demonstrated an improved accuracy and accelerated convergence of the model.

### 2.2.2. Cross-material and cross-process transfer

The versatility of knowledge transfer extends beyond dataset enhancement to enable knowledge transfer across different materials and manufacturing processes. Pandiyan et al. [63] explored this by training deep learning models on acoustic emissions from LPBF of stainless steel and successfully transferring the learned features to monitor LPBF of bronze. This cross-material transferability showcases the potential to generalize models across different materials, significantly reducing the model training time and the need for extensive new data collection for each material type. Similarly, Shin et al. [64] employed multisource knowledge transfer to detect balling defects in wire arc additive manufacturing (WAAM), by extracting features from multiple materials and finetuning the model for specific anomaly detection. Li et al. [65] conducted knowledge transfer from 316L stainless steel to TC4 titanium alloy for LPBF quality prediction based on a multisensor dataset, including layer-wise images, photodiode signals, and acoustic emission signals.

Knowledge transfer also facilitates the bridging differences in process maps, adapting models to handle variations in process parameters and environmental conditions effectively [66]. For example, Pandiyan et al. [67] employed unsupervised domain adaptation techniques to manage shifts in data distribution caused by different process parameter spaces in LPBF. Furthermore, knowledge transfer from one process to another is possible. Safdar et al. [68] proposed a structured framework for transferring data-driven knowledge between metal AM processes, such as from LPBF to LDED. The authors presented a three-step framework that supports the systematic transfer of knowledge at various levels, including data representation and model parameters, which enables efficient cross-process adaptations.

Overall, knowledge transfer offers substantial benefits in addressing the challenges of data scarcity and the high cost of data collection in LAM. By pretraining models on public domain datasets, this approach enables high defect detection accuracy with smaller, domain-specific datasets. In addition, it facilitates the transfer of knowledge from one material to another and between similar processes, such as from LPBF to LDED, both of which are fusion-based metal AM techniques. Despite these advancements, the potential for knowledge transfer from one sensing modality to another remains largely unexplored. This gap is significant because information from a superior modality, such as thermal images or coaxial melt pool images, could be transferred to more cost-effective modalities such as acoustic signals.

## 2.3. Cross-modality knowledge transfer

Cross-modality knowledge transfer leverages different data modalities to enrich learning models, overcoming limitations specific to one type of data by integrating another. This subsection systematically explores how this concept has been applied across various fields, enhancing model effectiveness and data utilization.

Athanasiadis et al. [69] tackled the challenge of emotion recognition, traditionally limited by the high costs and ambiguities of audio data annotation. By employing conditional semi-supervised generative adversarial networks, they fused facial expressions from images with audio signals to predict emotions more accurately. This cross-modality approach leveraged the rich, annotated visual data to compensate for the sparse audio annotations, demonstrating improved classification performance and the potential of visual data to enhance audio-based models. Similarly, Zhang et al. [70] introduced a combination of CMKT with semi-supervised learning. They used visual features to enhance the diversity and reliability of pseudo-labels in audio data, increasing the robustness and accuracy of speech (audio) model. This approach addressed the scarcity of labeled speech datasets and also mitigated the noise. Planamente et al. [71] addressed first-person action recognition problem, where environmental variability poses significant challenges. They developed an audio-visual loss that aligns the norm magnitudes of features from both modalities, enhancing the model's ability to generalize across different domains. Zhang et al. [72] explored activity recognition across visual and audio domains affected by changes such as scenery or camera viewpoint. They proposed an audio-adaptive encoder to adjust visual features based on audio features, reducing domain-specific biases and enhancing cross-domain applicability. This strategy highlighted the stabilizing effect of audio data on visual feature representations, especially in varied recording conditions. Cangea et al. [73] proposed a set of deep learning architectures for audio-visual classification that facilitates early integration of cross-modality data flows. This method harnessed the correlations between audio and visual data, enhancing the interpretability and performance of the resulting models. Furthermore, Guo et al. [74] introduced a deep visual-audio network for cross-modality retrieval between speech and remote sensing images, providing a solution to the inefficiencies of traditional text-based retrieval methods. By establishing direct associations between image features and audio features, this approach offered a faster and more intuitive way for retrieving images using natural language inputs.

These studies illustrate the effectiveness of CMKT in enhancing model performance, reducing reliance on large, labeled datasets, and improving generalization across varied conditions. While these applications have not yet been explored within the context of process monitoring in LAM, the CMKT principles could potentially be adapted to this field. For example, AV-HuBERT is a cross-modality learning method that extracts audio-visual speech representations from synchronized mouth images and audio recordings during speeches [75]. By exploiting the correlations between these two modalities, the learned representations can significantly improve the performance of lip-reading models, which rely solely on visual data for predictions. In AM, CMKT could mitigate the disadvantages of costly multimodal monitoring setups by removing microphones during the prediction phase, while learning salient audio-visual representations that enrich the useful information extracted solely from melt pool images. This untapped potential forms the foundation for investigating cross-modality domain adaptation in this study, aiming to enhance the adaptability and efficiency of monitoring systems in LAM.

## 3. Methodology

This section elaborates on the experiment setting and the proposed CMKT methods. The first subsection lays out the experiments that generated the multimodal dataset to demonstrate the effectiveness of the proposed CMKT methods. The second subsection illustrates the theories of the proposed CMKT methods, including semantic alignment, fully supervised mapping, and semi-supervised mapping.

### 3.1. Experiments and dataset descriptions

Figure 2 illustrates an LDED system along with audio-visual sensor setups, presented both schematically and in an actual photograph. The LDED system features a coaxial powder-feeding nozzle and a 1070 nm laser beam, which melts the metal powder materials for layer-by-layer deposition [76]. As the nozzle moves in the feed direction, the deposited material rapidly solidifies in the molten pool area. Two sensors are employed to monitor the dynamics of the melt pool: (1) a coaxial charge-coupled device (CCD) camera, operating at an acquisition frequency of 30 Hz, and (2) a microphone positioned approximately 20 cm from the process zone, with a sampling rate of 44100 Hz. Both sensors are interfaced with a personal computer running Ubuntu Operating System and an in-house developed software platform utilizing robot operating system (ROS). Audio and visual data were captured simultaneously during the experiments through this software, as detailed in our previous studies [50–52].

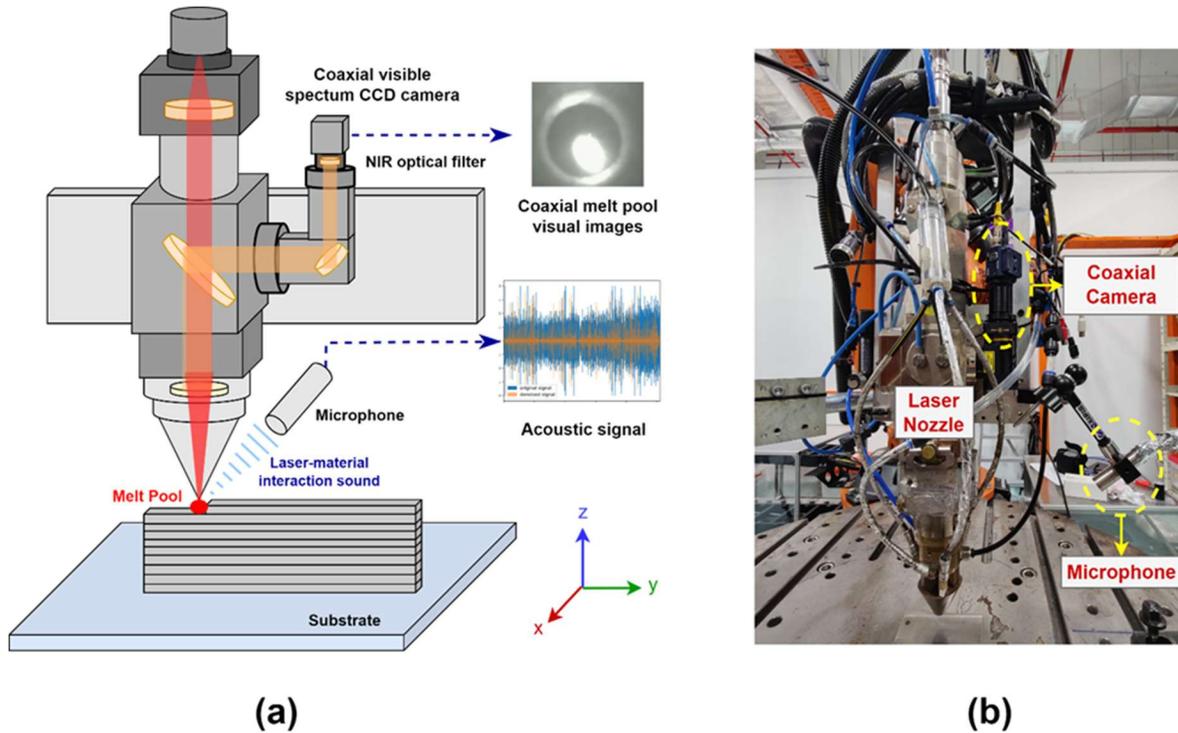

**Figure 2:** LDED system with coaxial melt pool camera and acoustic sensor setups: (a) schematic illustration and (b) experiment photo.

The audio-visual multimodal dataset was collected during the LDED process of single-bead walls with maraging steel C300 powder. Each single-bead wall consists of 50 layers. Table 1 presents the process parameters employed during the experiments. In this study, we opted not to introduce defects deliberately by using suboptimal parameters. Instead, parameters were pre-optimized to achieve near fully-dense quality, suitable for multi-layer, multi-track industrial production. However, due to limited heat transfer capacities, localized heat accumulation was exacerbated during the fabrication of single-bead wall samples. This led to rapid heating and cooling cycles, resulting in the gradual emergence of defects such as cracks and keyhole pores during the deposition. Following several deposited layers, the printing process showed a transition from defect-free to defective areas, changing the labels of the data points from these regions to "defective." Consequently, the dataset was segmented into three labeled categories: defective, defect-free, and laser-off. As single-bead wall samples were deposited, the dwell time between each layer was incrementally extended (from 0 to 5 to 10 seconds) to mitigate localized heat accumulation and postpone defect formation by allowing additional cooling time. As a result, the occurrence of defects varied across samples, appearing at different layers. Samples deposited with longer dwell times demonstrated fewer defects. The single-bead walls were wire-cut to expose their vertical cross-sections, allowing the identification of defect locations across layers and their spatial coordinates. Optical microscope (OM) images were captured to identify the presence and location of cracks and keyhole pores within the components, serving as the physical labels to guide defect prediction. Audio-visual signals captured during the deposition were synchronized with the robot toolpath coordinates using in-house ROS-based software. This synchronization directly correlated the observed defects in OM images and the audio-visual signals recorded along the deposition path, streamlining the labeling process. Further information on the experimental procedures and datasets is available in our prior publication [51]. The example dataset has also been made publicly accessible for download by interested readers [77].

**Table 1: LDED process parameters and setup information for audio-visual data collection experiments.**

| Parameters | Values |
| --- | --- |
| Part geometry | Single-bead wall structure |
| Laser power (kW) | 2.3-2.5 |
| Speed (mm/s) | 25-27.5 |
| Dwell time between the layers (s) | [0, 5, 10] |
| Laser beam diameter (mm) | 2 |
| Powder flow rate (g/min) | 12 |
| Energy density (kW·s /mm) | 0.92 |
| Hatch space (mm) | 1 |
| Layer thickness (mm) | 0.85 |
| Stand-off distance (mm) | 12 |
| Material | Maraging Steel C300 |
| Types of defects generated | Cracks, keyhole pores |

The LDED experiment generated a synchronized and registered multimodal dataset consisting of a visual dataset in JPG format and an audio dataset in WAV format as illustrated in [51]. Data handling techniques were applied to prepare and preprocess the dataset for subsequent knowledge transfer and prediction tasks (Figure 3). The visual dataset had 4345 melt pool images, each having three channels and 480 × 480 pixels in each channel. To reduce the dimensionality while retaining most of the information, each melt pool image was grayscaled and resized from 480 × 480 pixels to 80 × 80 pixels, resulting in a matrix shape of 4345 × 80 × 80. The audio dataset was comprised of 6.387 million audio frames in 1 channel, thus having a matrix size of 6.387e6 × 1. The audio data was segregated into audio snippets of 33.3 milliseconds corresponding to the acquisition frequency of the visual data. This transformation resulted in an audio dataset with a matrix shape of 4345 × 1470. Each audio snippet stored the audio signals collected between two adjacent melt pool images. The same label was shared between an audio snippet and the melt pool image captured at the beginning of the snippet.

One of the challenges of multimodal and cross-modality learning is aligning modalities with different data types. The two datasets differ in shape because the visual dataset contains image data, while the audio dataset contains time-series data. In the ML domain, researchers typically deploy appropriate encoders that map one modality to the other or that map two modalities into a shared latent space [20]. For example, to obtain audio-visual representations, Shi et al. [75] mapped mouth images and speech recordings into a latent space using ResNet and a feedforward network, respectively. However, training multiple encoders requires a large multimodal dataset, which is rarely available in ML-based AM monitoring [14].

Alternatively, two modalities can be aligned using data preparation based on engineering domain knowledge. In this work, each audio snippet was converted to a spectrogram, which is proven efficient for extracting salient audio features [78]. The horizontal axis, vertical axis, and color scale of a spectrogram represent the time, frequency, and magnitude of the audio, respectively. The conversion parameters were adjusted to output spectrograms of 80 × 80 pixels, ensuring that the shapes of the two modalities were equalized. The frequency axis of the spectrogram follows a linear scale rather than a logarithmic scale because the latter compresses high-frequency ranges, which contain useful information for AM defect detection [41]. Consequently, each window in the spectrograms represents 0.0416 milliseconds and each bin corresponds to a frequency range of 275 Hz. Following this, the spectrograms were converted to grayscale images for dimensionality reduction. This conversion ensured that both the visual and audio input data had the same shape, allowing them to be fed into a shared network, which reduces model complexity. The experiment yielded 4345 melt pool images and audio snippets, which were divided into a training set, a validation set, and a test set at a ratio of 8:1:1. The data size of each sub-dataset is presented in Table 2. It is difficult to differentiate between defect-free and defective examples through visual inspections of the raw audio data, spectrograms, and melt pool images, which highlights the necessity of employing ML techniques.

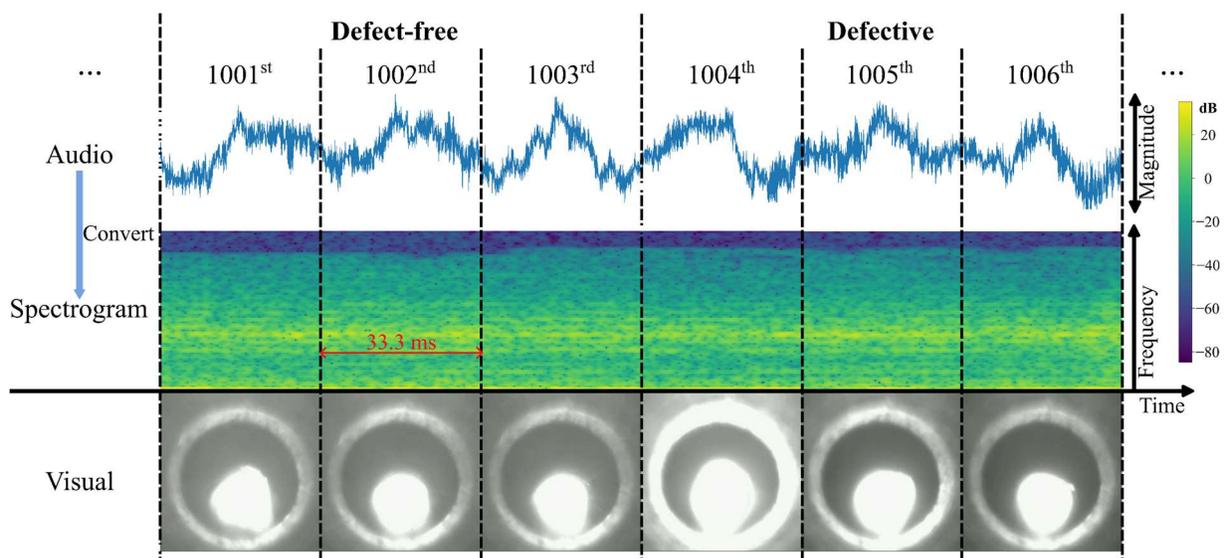

**Figure 3: Synchronized and registered audio data (converted to spectrograms) and visual data (melt pool images).**

**Table 2: Train-validation-test split of the dataset.**

| Datasets | Defect-free | Defective |
| --- | --- | --- |
| Training | 859 | 2617 |
| Validation | 112 | 322 |
| Test | 114 | 321 |

## 3.2. Cross-modality knowledge transfer

Knowledge transfer, including transfer learning and domain adaptation, involves knowledge sharing between a source and a target [78]. In transfer learning, the tasks (i.e., output variables) of the source and target are different, while the input domain remains the same. CMKT belongs to domain adaptation, where the input domains are different between the source and target, while the task remains the same ($\mathcal{Y}_V = \mathcal{Y}_A$). The multimodal AM in-situ monitoring system illustrated in Figure 2 involves a visual dataset ($D_V$) and an audio dataset ($D_A$). The visual dataset ($D_V = \{(x_{V,1}, y_{V,1}), \ldots, (x_{V,n_V}, y_{V,n_V})\}$) with a data size of $n_V$ is collected using the coaxial camera, where $x_{V,i} \in \mathcal{X}_V$ and $y_{V,i} \in Y_V$ are the i$^{th}$ melt pool image and the corresponding label, respectively. The audio dataset ($D_A = \{(x_{A,1}, y_{A,1}), \ldots, (x_{A,n_A}, y_{A,n_A})\}$) with a data size of $n_A$ is collected using the microphone, where $x_{A,i} \in \mathcal{X}_A$ and $y_{A,i} \in Y_A$ are the i$^{th}$ spectrogram and the corresponding label, respectively. The input domains in this setting have different feature spaces ($\mathcal{X}_V \neq \mathcal{X}_A$) and marginal probability distributions ($P(X_V) \neq P(X_A)$). In CMKT, both $D_V$ and $D_A$ can be used to train the model, while only the input of one modality ($X_V$ or $X_A$) is needed for making predictions. Knowledge can be transferred from the visual modality to the audio modality, or from the audio modality to the visual modality. The remaining modality at the prediction phase is the target modality (Figure 1b). This paper proposes three CMKT methods for AM in-situ monitoring: semantic alignment, fully supervised mapping, and semi-supervised mapping.

The three proposed CMKT methods are designed to extract salient representations shared between modalities by exploiting the correlations between the visual and audio data, thereby minimizing noises that are usually modal-specific. In this multimodal monitoring system, both the camera and microphone are deployed to capture information about the same physical phenomena: melting and solidification of metal powders. Similar to lip reading, where mouth shapes and speech recordings are highly correlated [75], melt pool shapes and audio recordings of the melting process exhibit highly correlated signatures useful for defect detection [57,58,79]. However, irrelevant information, such as sensor and environmental noises, can be introduced into the dataset during multimodal data acquisition. For example, random noise and flare may arise from optical imaging, obscuring the morphological characteristics of melt pools. CMKT can address these challenges by leveraging the modality-specific nature of such noises and focusing on features shared between both modalities.

### 3.2.1. Semantic alignment

The first proposed method utilizes classification and contrastive semantic alignment (CCSA) [80]. Semantic alignment conducts knowledge transfer by aligning the source and the target distributions in an encoded space [80]. The CCSA architecture adapted for CMKT is presented in Figure 4a, which illustrates the model structure, loss terms, and forward and backward propagations. The model structure consists of a convolutional encoder ($G_e$) and a task classifier ($G_t$), which are both shared between the visual and audio modalities. $G_e$ extracts salient representations from visual and audio input data utilizing concatenated convolution layers, pooling layers, and activation functions. An encoded space, constructed by $G_e$, semantically aligns the visual and audio distributions using a contrastive semantic alignment loss ($L_{CSA}$):

$$L_{CSA} = L_{SA} + L_S, \qquad (1)$$

where $L_{SA}$ and $L_S$ are semantic alignment loss and separation loss, respectively. $L_{SA}$ encourages $G_e$ to aggregate samples with the same labels from different modalities in the encoded space:

$$L_{SA}(G_e) = \sum_{a=1}^{C} d\left(p(G_e(X_V^a)), p(G_e(X_A^a))\right), \qquad (2)$$

where C is the number of classes, d is a distance metric between two distributions, $X_V^a = X_V|\{Y = a\}$ and $X_A^a = X_A|\{Y = a\}$. In other words, $L_{SA}$ brings visual and audio defective groups closer and visual and audio defect-free groups closer, forcing the extraction of features shared between two modalities (Figure 4b). $L_S$ penalizes the similarity between samples with different labels from different modalities in the encoded space:

$$L_S(G_e) = \sum_{a,b|a \neq b} k\left(p(G_e(X_V^a)), p(G_e(X_A^b))\right), \qquad (3)$$

where $k$ is a similarity metric. To learn a shared encoded space for both modalities, $G_e$ utilizes a Siamese network shared between the visual and audio data [81]. In other words, $L_S$ increases the distance between visual defective and audio defect-free groups and the distance between visual defect-free and audio defective groups, forcing the extraction of task-relevant and modal-specific features (Figure 4b). A shared network also reduces computational complexity. Based on the salient representations extracted by $G_e$, a shared task classifier ($G_t$) classifies whether the sample belongs to the defective or defect-free class. At each training epoch, $G_t$ is updated using the mean classification loss ($L_C$), which is the mean of the visual and audio task classification losses ($L_V$ and $L_A$):

$$L_C(G_e \circ G_t) = [L_V(G_e \circ G_t) + L_A(G_e \circ G_t)]/2. \qquad (4)$$

In this paper, $L_V$ and $L_A$ are calculated using binary cross entropy, which encourages the model to split the visual defective and defect-free groups and to split the audio defective and defect-free groups (Figure 4b). Following the update of $G_t$, $G_e$ is updated using the CCSA loss:

$$L_{CCSA}(G_e \circ G_t) = (1 - \gamma)(L_{SA}(G_e) + L_S(G_e)) + \gamma L_C(G_e \circ G_t), \qquad (5)$$

where $\gamma$ is a trade-off factor between $L_C$ and $L_{CSA}$. In this way, the two distributions must be well aligned in the encoded space to obtain accurate classifications because $G_t$ is also shared between the two modalities. In other words, the representations of the two modalities must be similar for securing accurate classifications using the same task classifier.

In a real-life setting, manufacturing datasets almost always face data scarcity. However, both $L_{SA}$ and $L_S$ calculates the distributions of the encoded visual and audio data, which demands a large data size. Motiian et al. [80] suggests approximating the distance metric ($d$) by computing the average pairwise distances between the encoded samples:

$$d\left(p(G_e(X_V^a)), p(G_e(X_A^a))\right) = \sum_{i,j} d\left(G_e(x_V^i), G_e(x_A^j)\right), \qquad (6)$$

where $y_V^i = y_A^j = a$. Likewise, the similarity metric ($k$) can also be approximated by pairwise distances:

$$k\left(p(G_e(X_V^a)), p(G_e(X_A^b))\right) = \sum_{i,j} k\left(G_e(x_V^i), G_e(x_A^j)\right), \qquad (7)$$

where $y_V^i = a \neq y_A^j = b$. It can be further assumed that:

$$d\left(G_e(x_V^i), G_e(x_A^j)\right) = \|G_e(x_V^i) - G_e(x_A^j)\|^2/2, \tag{8}$$

$$k\left(G_e(x_V^i), G_e(x_A^j)\right) = \max\left(0, m - \|G_e(x_V^i) - G_e(x_A^j)\|\right)^2/2, \tag{9}$$

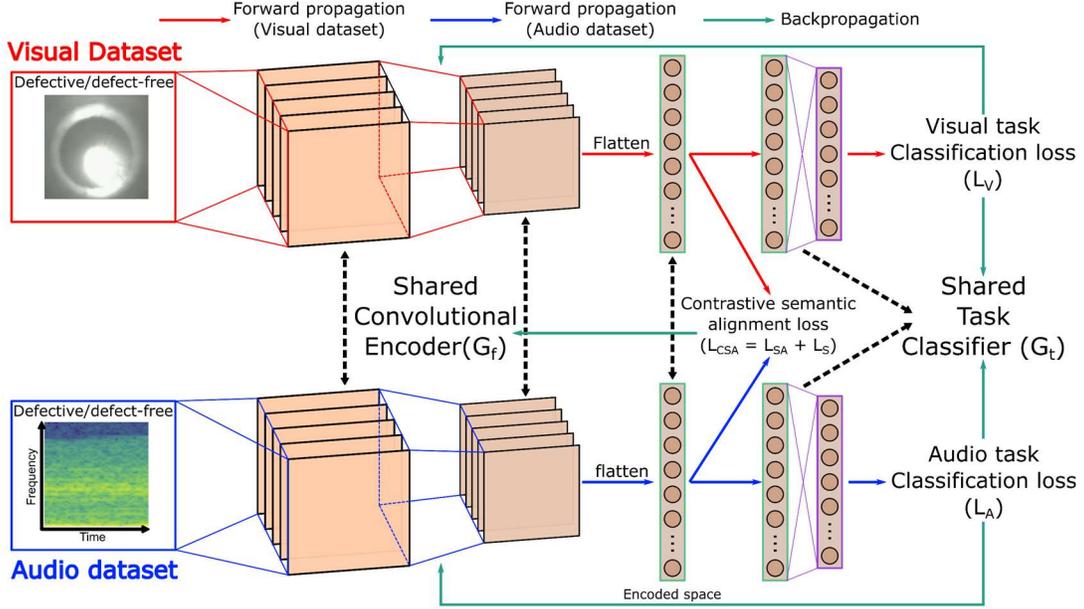

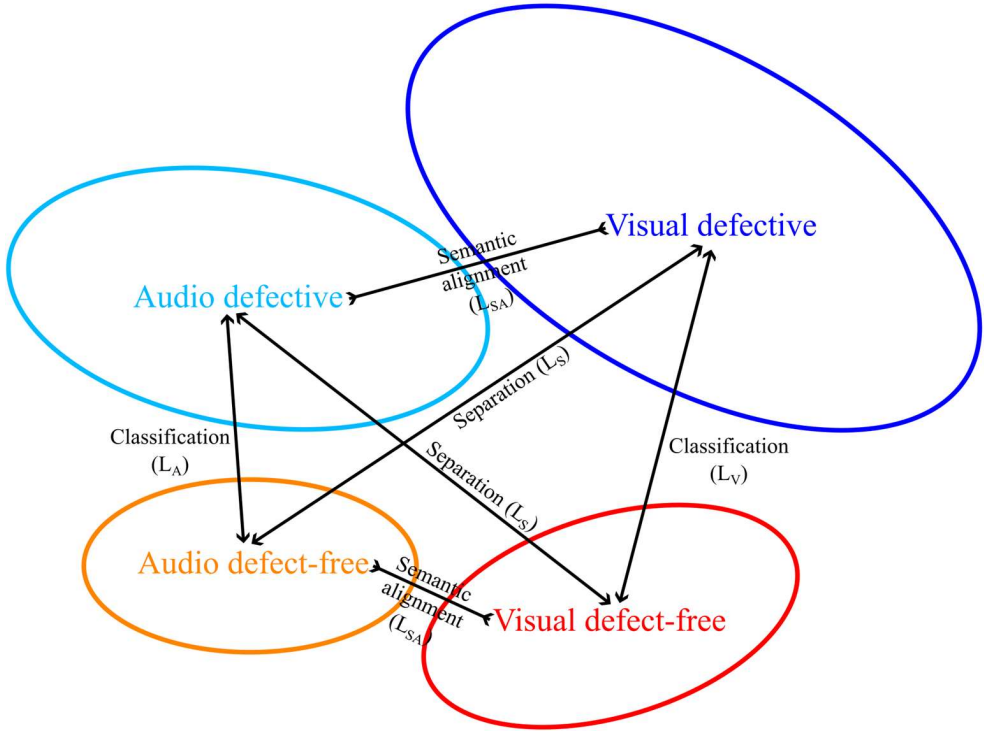

**Figure 4: Schematics of the proposed cross-modality knowledge transfer (CMKT) based on contrastive and classification semantic alignment (CCSA): a) model architecture and b) impact of the loss terms illustrated in the encoded space.**

where $\|\cdot\|$ denotes the Euclidean norm and $m$ is the margin that defines the separability in the encoded space. If knowledge is transferable between the two modalities, a trained model should be able to extract salient representations and make accurate classifications for both modalities. The model only needs the input data collected from visual or audio sensors to make predictions.

### 3.2.2. Cross-modality mapping

The second proposed strategy conducts knowledge transfer by learning a mapping between two modalities to derive the salient representations of one modality from the other modality (Figure 5). Unlike semantic alignment, mapping requires defining a source and a target. The source modality could be removed during the prediction phase if its salient features are derivable from the target modality. This strategy usually involves multiple phases: Phase 1 utilizes feature learning to extract salient representations, Phase 2 learns a mapping between two modalities, and Phase 3 finetunes the mapping to adapt to the prediction task. The mapping strategy can be categorized into fully supervised (Figure 5a) and semi-supervised (Figure 5b) mappings that extract salient representations using supervised and unsupervised learning approaches, respectively.

*Fully supervised mapping*

Figure 5a illustrates how fully supervised mapping CMKT can be conducted from the visual domain to the audio domain in three phases. The first phase trains a classification model ($G_V$) purely based on the visual dataset. The classification loss used to train $G_V$ can be expressed as $L_V(G_V(X_V), Y_V)$. The hidden features ($F_h$) that are representative of the classification task are extracted from the last hidden layer of $G_V$. The second phase utilizes $F_h$ to train a mapping, $G_{A1}$, and guide the feature extraction from audio samples, which conducts knowledge transfer from the visual modality to the audio modality. $G_{A1}$ is trained to minimize $L_{A1}(G_{A1}(X_A), F_h)$. In the third phase, $G_{A1}$ is fixed to extract $F_h$ from audio samples, followed by additional hidden layers ($G_{A2}$) to make a classification based on $F_h$. The classification loss used to train $G_{A2}$ can be expressed as $L_{A2}(G_{A2}(F_h), Y_A)$. This method can also transfer knowledge from the audio to the visual modality by switching the positions of the two modalities in Figure 5a. In this way, $G_{A1}$ learns to derive the hidden feature of the audio modality from the visual modality. However, the features extracted from the visual modality are usually more effective for defect detection than the features from the audio modality. Thus, knowledge transfer from the audio to the visual modality might lead to compromised classification performance.

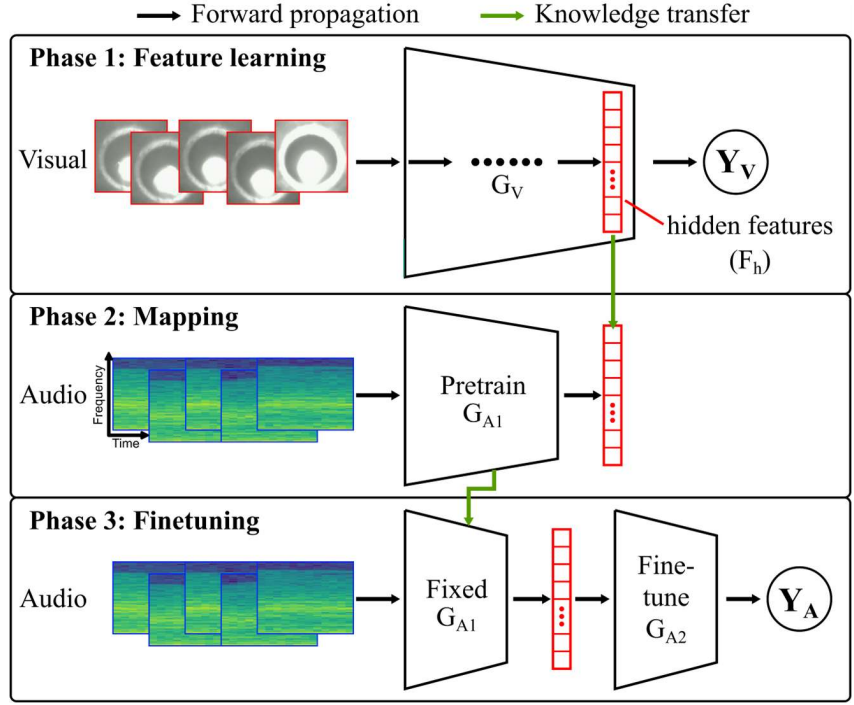

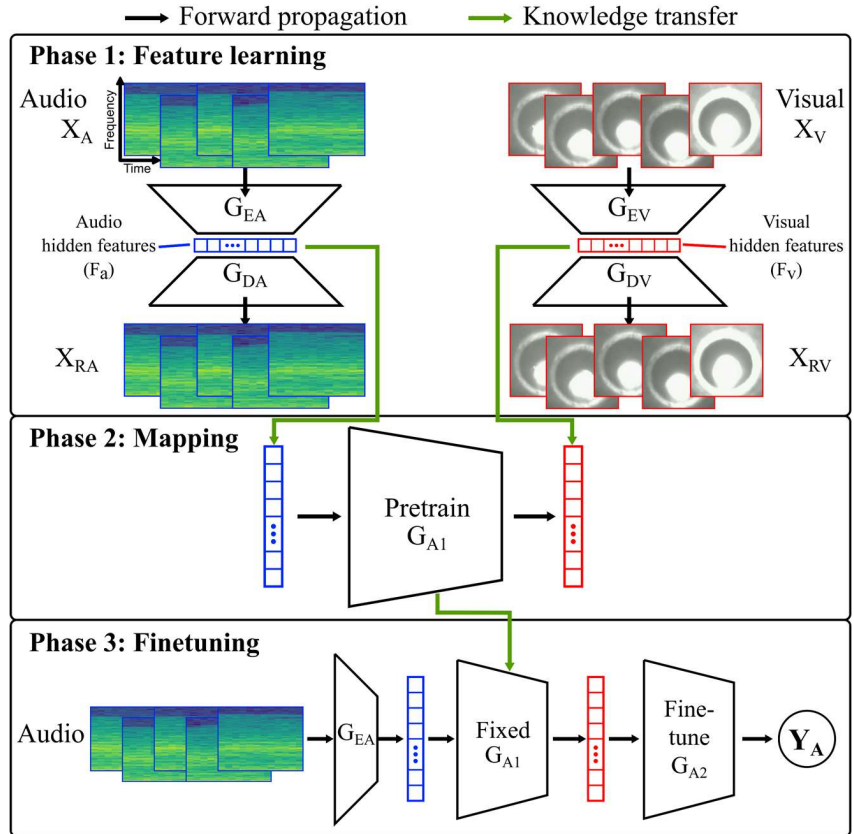

**Figure 5:** Schematics of the cross-modality mapping methods: (a) fully supervised mapping and (b) semi-supervised mapping.

*Semi-supervised mapping*

Figure 5b illustrates how semi-supervised mapping CMKT can be conducted from the visual domain to the audio domain in three phases. The first phase separately trains two autoencoders: $G_V$ consisting of $G_{EV}$ and $G_{DV}$ based on $X_V$ and $G_A$ consisting of $G_{EA}$ and $G_{DA}$ based on $X_A$. The reconstruction loss used to train $G_A$ and $G_V$ can be expressed as $L_A(X_A, X_{RA})$ and $L_V(X_V, X_{RV})$ respectively where subscript 'R' represents reconstructed inputs. The hidden features ($F_a$ and $F_v$) that are representative of the input modalities are extracted from the bottleneck layers of $G_A$ and $G_V$, respectively. The second phase utilizes $F_a$ and $F_v$ to train a model, $G_{A1}$, and guide the feature mapping from audio to visual hidden features, which conducts knowledge transfer from the visual modality to the audio modality. $G_{A1}$ is trained to minimize $L_{A1}(F_a, F_v)$. In the third phase, $G_A$ and $G_{A1}$ are fixed to extract $F_v$ from audio samples, followed by an additional model ($G_{A2}$) to make a classification based on $F_v$. The classification loss used to train $G_{A2}$ can be expressed as $L_{A2}(G_{A2}(G_{A1}(G_{EA}(X_A))), Y_A)$. By switching the direction of the forward propagation at Phase 2 of Figure 5b, this method can also transfer knowledge from the audio modality to the visual modality. In this setting, $G_{A1}$ learns to map the features from the visual modality to the audio modality. However, as highlighted by the related research, the features extracted from the visual modality are usually more information-rich than the features from the audio modality and thus more effective for defect detection [15,82].

## 4. Results

This section demonstrates the training and evaluation of the proposed CMKT methods. As a comparison, three multimodal fusion methods, including data-level fusion, feature-level fusion, and decision-level fusion, were implemented and evaluated in this section. Besides, single-modal methods (i.e., visual-data-only and audio-data-only) were implemented as the baselines. Hyperparameter searches were conducted to select the optimal models and investigate the performance of each method. Finally, the methods were evaluated using three metrics namely accuracy, area under the receiver operating characteristics curve (AUC-ROC), and balanced accuracy.

### 4.1. Model training and hyperparameter search

Comprehensive hyperparameter searches were performed to establish a fair comparison and to obtain the optimal performance of the CMKT and multimodal fusion methods. For each method, 300 candidate models were trained using the training set and evaluated using the validation set. For example, Table 3 presents the model settings and hyperparameter ranges of an extensive hyperparameter search conducted for the semantic alignment CMKT method. This hyperparameter search explored a wide range of learning rates, numbers of convolution layers, numbers of filters, kernel sizes, numbers of fully connected layers, numbers of neurons, and dropout rates. According to the literature and preceding training results, a separability margin ($m$) of 1, a trade-off factor ($\gamma$) of 0.5, and 1200 training epochs were selected to ensure convergence. The training process also deployed class weighting to address class imbalance in the dataset. The hyperparameter search employed Bayesian optimization that maximized the validation accuracy to obtain the optimal hyperparameters. We conducted the hyperparameter search using Optuna [83].

**Table 3: Model settings and hyperparameter search of the semantic alignment CMKT method.**

| Fixed hyperparameter/setting | Value/method |
|---|---|
| Activation | ReLU |
| Optimizer | Adam |
| $L_V$ and $L_A$ | Binary cross-entropy |
| Class weighting | defective/defect-free=1/3 |
| $L_{CCSA}$ | $(1-\gamma)(L_{SA}+L_S)+\gamma(L_V+L_A)/2$ |
| Input and output layer dimensions | $80 \times 80$ and 1 |
| Separability margin ($m$) | 1 |
| Trade-off factor ($\gamma$) | 0.5 |
| Learning rate | $(x \in \mathbb{R} | 1E^{-6} \leq x \leq 1E^{-3})$ |
| Weight decay | $(x \in \mathbb{R} | 1E^{-7} \leq x \leq 1E^{-3})$ |
| Number of epochs | 1200 |
| **Convolutional encoder ($G_e$)** | |
| **Tuned hyperparameters** | **Range** |
| Number of convolution layers | $(x \in \mathbb{N} | 3 \leq x \leq 5)$ |
| Number of filters at each convolution layer | $(x \in \mathbb{N} | 16 \leq x \leq 48)$ |
| Kernel size at each convolution layer | $(x \in \mathbb{N} | 2 \leq x \leq 4)$ |
| **Task classifier ($G_t$)** | |
| **Tuned hyperparameters** | **Range** |
| Number of fully connected layers | $(x \in \mathbb{N} | 1 \leq x \leq 3)$ |
| Number of neurons at each hidden layer | $(x \in \mathbb{N} | 32 \leq x \leq 360)$ |
| Dropout rate | $(x \in \mathbb{R} | 0.01 \leq x \leq 0.1)$ |

The top 50 models with the highest validation accuracies were selected and trained again to evaluate their test performance and reproducibility. The structure of the optimal model that achieved the highest validation accuracy was recorded in Table 4. It was a deep CNN model consisting of 5 convolutional modules in $G_e$ and 3 linear modules in $G_t$. The model incorporated regularization techniques including weight decay and dropout to mitigate overfitting during training. The model training and hyperparameter searches of the other implemented methods followed the same procedures as illustrated above. The structures of the optimal fully supervised mapping and semi-supervised mapping models are shown in the appendix (Table A.1 and Table A.2).

To investigate the effects of semantic alignment CMKT, the training process is visualized in Figure 6. Figure 6a visualizes the encoded space to illustrate the relationships among the audio defect-free, audio defective, visual defect-free, and visual defective groups. After every 150[th] epoch, the test examples were mapped into the encoded space and subsequently reduced to a two-dimensional space using t-distributed stochastic neighbor embedding (t-SNE). Initially, the visual defect-free examples were clustered with the visual defective examples, and the audio defect-free examples were clustered with the audio defective examples. Through the training process, the visual defect-free group gradually moved away from the visual defective group and being aligned with the audio defect-free group. Likewise, the audio defective group moved away from the audio defect-free group and shifted toward the visual defective group. At the 1200[th] epoch, defective examples formed a cluster at the top and the defect-free examples formed a cluster at the bottom in the visualization of the encoded space. A small portion of the visual defect-free examples stayed at the top but away from the visual defective group.

**Table 4: The optimal hyperparameters of the semantic alignment CMKT model.**

| Common settings | |
|---|---|
| Initial learning rate | 0.0007322092 |
| Weight decay | 0.0005568733 |
| **Convolutional encoder ($G_e$)** | |
| *Layer* | *Hyperparameters* |
| 1st Convolutional module | Conv2d (31 channels, kernel size=2, stride=1, padding= 'same'), ReLU(), MaxPool2d (kernel size=2, stride=1) |
| 2nd Convolutional module | Conv2d (31 channels, kernel size=2, stride=1, padding= 'same'), ReLU(), MaxPool2d (kernel size=1, stride=1) |
| 3rd Convolutional module | Conv2d (31 channels, kernel size=2, stride=1, padding= 'same'), ReLU(), MaxPool2d (kernel size=1, stride=1) |
| 4th Convolutional module | Conv2d (32 channels, kernel size=2, stride=1, padding= 'same'), ReLU(), MaxPool2d (kernel size=2, stride=1) |
| 5th Convolutional module | Conv2d (32 channels, kernel size=2, stride=1, padding= 'same'), ReLU(), MaxPool2d (kernel size=2, stride=1) |
| Flatten | Flatten (32 × 10 × 10) |
| **Task classifiers ($G_t$)** | |
| *Layer* | *Hyperparameter/setting* |
| 1st Fully connected module | Linear (32 × 10 × 10, 198), ReLU(), Dropout (0.01821) |
| 2nd Fully connected module | Linear (198, 41), ReLU(), Dropout (0.01821) |
| 3rd Fully connected module | Linear (41, 1), sigmoid() |

To quantitatively evaluate the impact of semantic alignment in the encoded space, the maximum mean discrepancies (MMDs) between the four groups were computed (Figure 6b). MMD is a widely used non-parametric statistical test for comparing two probability distributions based on their data samples [84]. MMD works by embedding the samples into a reproducing kernel Hilbert space (RKHS) and computing the difference between the means of the two distributions in this space. For two distributions $P$ and $Q$, MMD is defined as:

$$MMD(P,Q) = \left\| \mathbb{E}_P[\phi(x_1)] - \mathbb{E}_Q[\phi(x_2)] \right\|_{\mathcal{H}}, \tag{10}$$

where $\mathcal{H}$ denotes the RKHS and $\phi(\bullet)$ is the function that maps the data into the RKHS. MMD is frequently applied in ML tasks such as domain adaptation to evaluate the similarity between datasets [85]. In this study, we computed the MMDs between the four groups: MMD between the audio defect-free and defective groups ($d_A$), MMD between the visual defect-free and defective groups ($d_V$), MMD between the visual and audio defect-free groups ($d_{VA,defect-free}$), and MMD between the visual and audio defective groups ($d_{VA,defective}$). At the 150th epoch, $d_V$ has reached 0.469 while $d_A$ was still smaller than 0.1, indicating the difficulty in classifying defect-free and defective examples based on the audio modality (Figure 6b). Meanwhile, $d_{VA,defect-free}$ (0.225) and $d_{VA,defective}$ (0.344) were much greater than $d_A$, which means that despite belonging to the same class, two groups from different modalities were far from each other in the encoded space.

Through semantic alignment CMKT, $d_A$ and $d_V$ rapidly increase to 0.437 and 0.762, respectively, at the $1200^{th}$ epoch, while $d_{VA,defect-free}$ and $d_{VA,defective}$ gradually decreased to 0.022 and 0.029, respectively. Figure 6b quantitatively shows that two groups belonging to the same class but from different modalities, though initially far apart, are drawn closer by semantic alignment CMKT. This occurs because semantic alignment encourages the model to extract shared representations between the two modalities, thereby amplifying the similarities between groups that belong to the same class.

The loss and accuracy curves can be found in Figure 6c. Both $L_{CSA}$ and $L_C$ rapidly descended in the first 400 epochs, then slowly converged at the $1200^{th}$ epoch. As a combination of $L_{CSA}$ and $L_C$, $L_{CCSA}$ followed the same trend as the two loss terms. A decreasing $L_{CSA}$ indicated that the source and target distributions were being mixed and that the examples of the same class were being aligned in the encoded space. A decreasing $L_C$ implied that $G_t$ was learning to classify defect-free and defect examples based on both the source and target modalities. Consequently, the training accuracy also increased at the same pace as $L_C$.

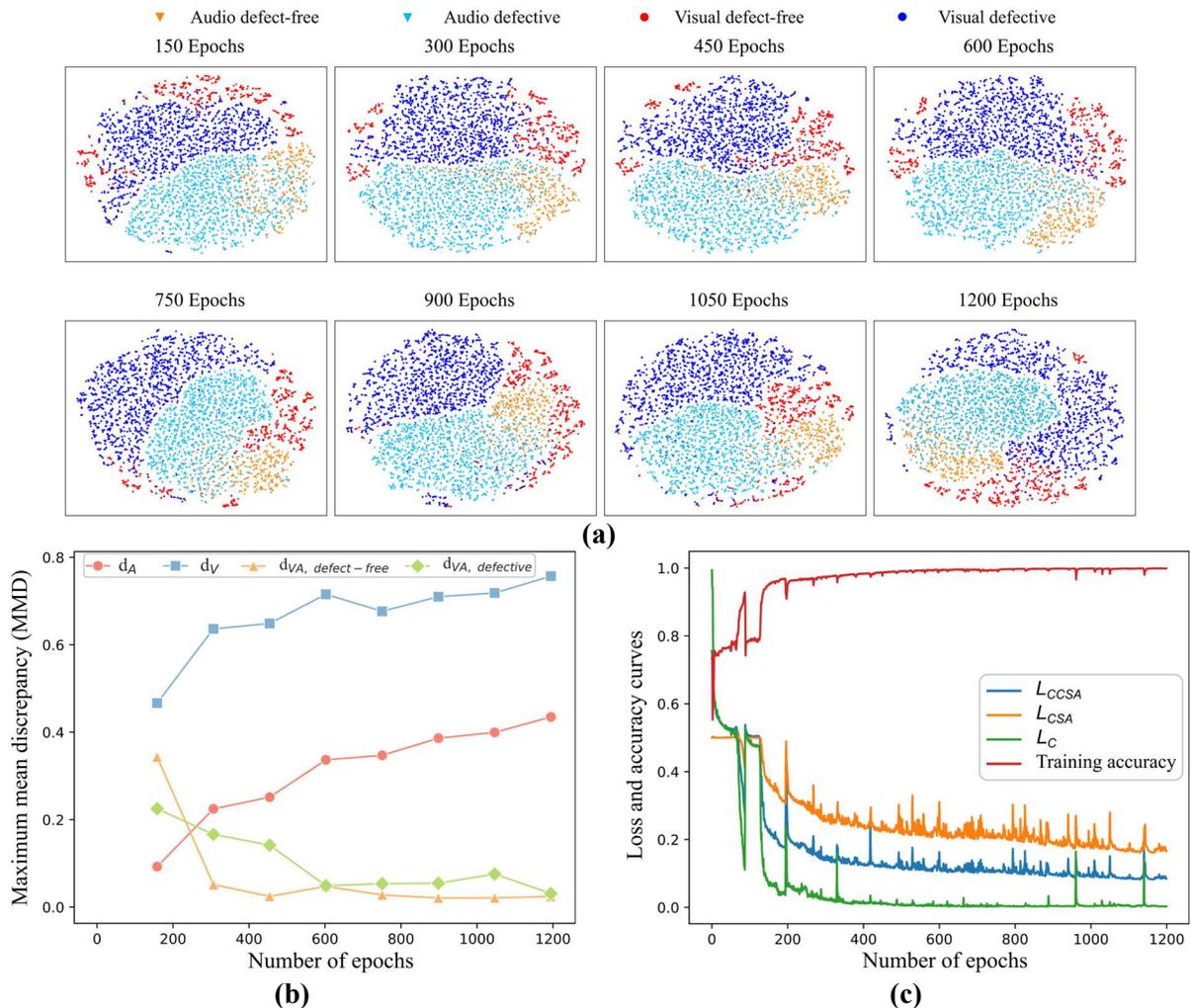

**Figure 6: Training process of the semantic alignment CMKT model with the highest validation accuracy: a) t-SNE visualization of the encoded space at every $150^{th}$ epoch b) maximum mean discrepancies; and c) loss and training accuracy curves.**

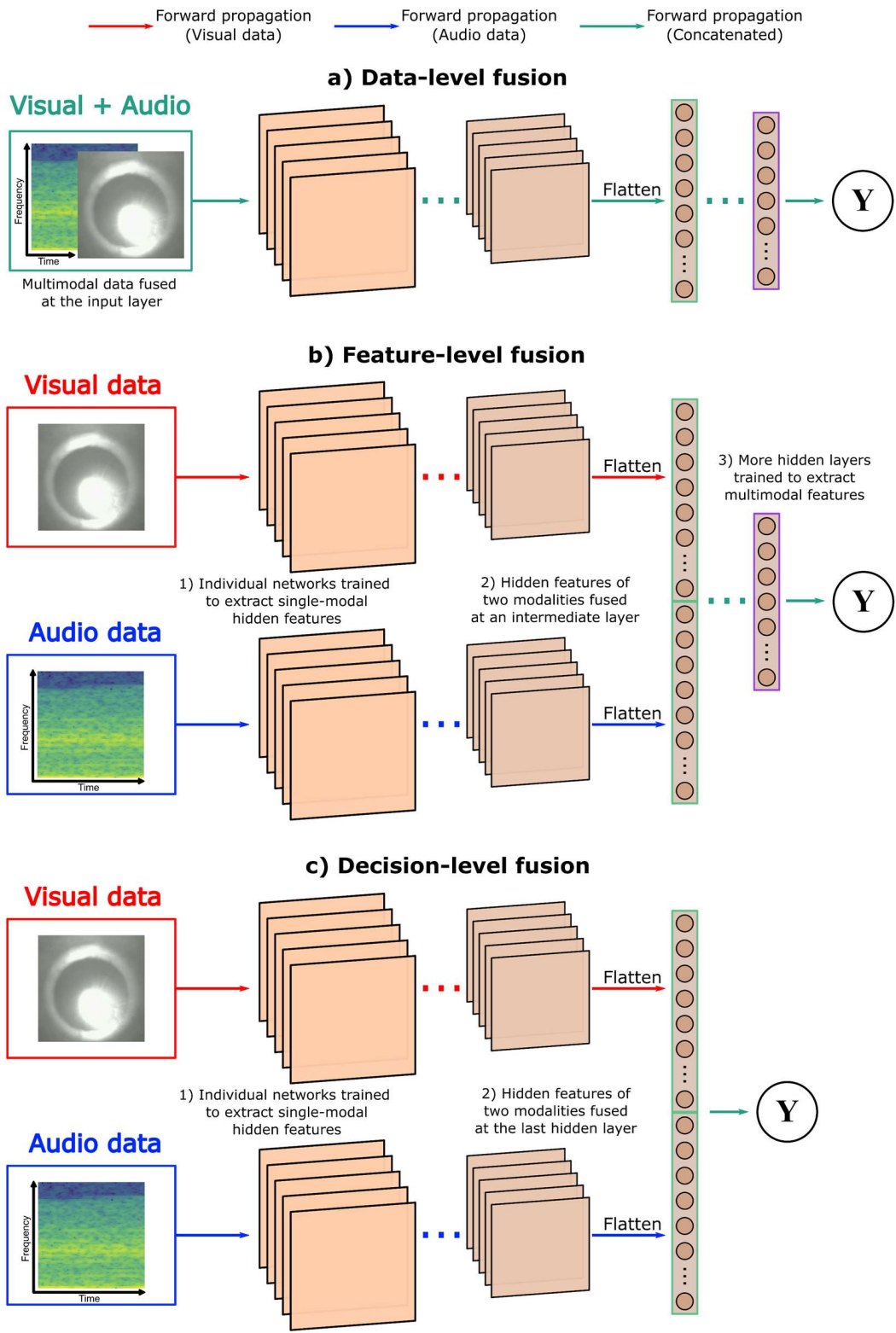

**Figure 7: Multimodal fusion (a) data-level fusion; (b) feature-level fusion; and (c) decision-level fusion.**

### 4.2. Multimodal fusion

Multimodal fusion methods were also implemented to compare with the proposed CMKT methods. Multimodal fusion integrates data from multiple modalities to derive rich multimodal representations for the modeling task. Figure 7 illustrates three common forms of multimodal fusion: data-level, feature-level, and decision-level fusion. Data-level fusion concatenates the input data of two modalities, and multimodal representations are extracted using a single model, as shown in Figure 7a. Feature-level fusion involves learning salient representations for each modality separately, then concatenating these representations and feeding them into subsequent hidden layers. The hidden layers are trained to extract multimodal representations and make predictions (Figure 7b). In decision-level fusion, representations are learned separately from each modality and concatenated, but predictions are made directly from the concatenated representations without additional hidden layers (Figure 7c).

### 4.3. Model evaluation

The top models selected from the hyperparameter search of each method were evaluated using the test set based on three metrics: accuracy, AUC-ROC, and balanced accuracy. These three metrics were calculated based on the classification results, including the true positive (TP), true negative (TN), false positive (FP), and false negative (FN). Accuracy is the ratio between the correct predictions and total predictions, which can be obtained using:

$$Accuracy = (TP + TN)/(TP + TN + FP + FN), \qquad (11)$$

AUC-ROC evaluates model robustness based on its ability to discriminate between different classes. The ROC curve of a model plots the true positive rate (TPR) against the false positive rate (FPR) at different classification thresholds. Therefore, a large AUC of the ROC curve indicates that the model secures high TPRs and low FPRs across a wide range of thresholds. Balanced accuracy is the arithmetic mean of TPR and true negative rate (TNR) ($Balanced\ accuracy = (TPR + TNR)/2$). It accounts for the data imbalance in the dataset because the model performance on both classes is equally weighted.

The test performance of the top 50 models of each method was exhibited in Figure 8 and Figure 9. The bar values, upper limits, and lower limits represent the average, optimal, and minimum scores of the top 50 models, respectively. Methods with the highest optimal scores demonstrate the best performance. Meanwhile, methods with high average scores are more robust, as the majority of their selected models consistently achieve high performance. Methods with small ranges (i.e., the difference between the upper and lower limits) can provide consistent performance because most of their selected models achieved similar performance.

As illustrated in Section 3.2.2, the mapping CMKT methods have two knowledge transfer directions: from the visual to the audio modality and from the audio to the visual modality. When knowledge is transferred from the audio to the visual modality, the prediction phase takes visual data as input and maps them to the audio modality. Thus, the CMKT of this direction was compared with the visual-data-only single-modal method (Figure 8). When knowledge is transferred from the visual to the audio modality, the prediction phase takes audio data as input and maps them to the visual modality. Thus, the CMKT of this direction was compared with the audio-data-only single-modal method (Figure 9). Semantic alignment does not map the input to the other modality but to the encoded space shared between the two modalities.

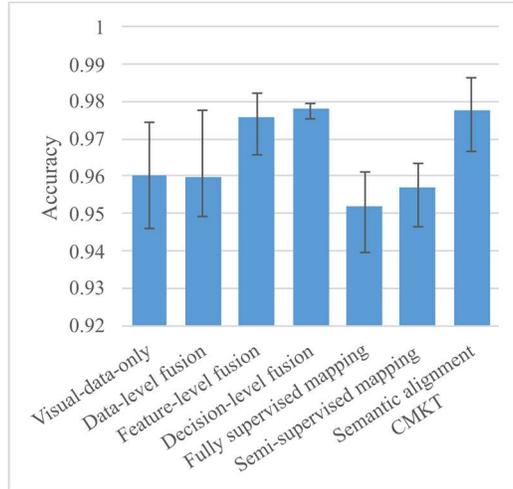

(a)

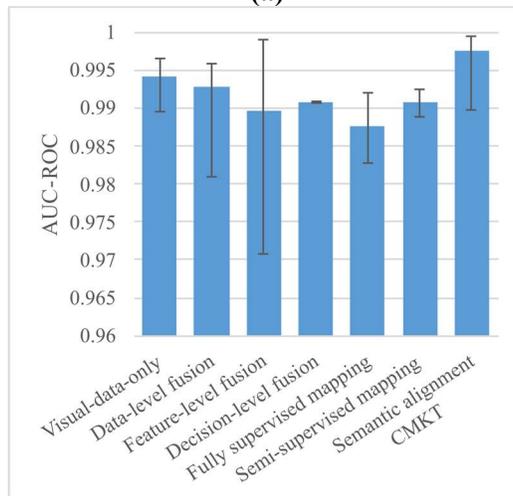

(b)

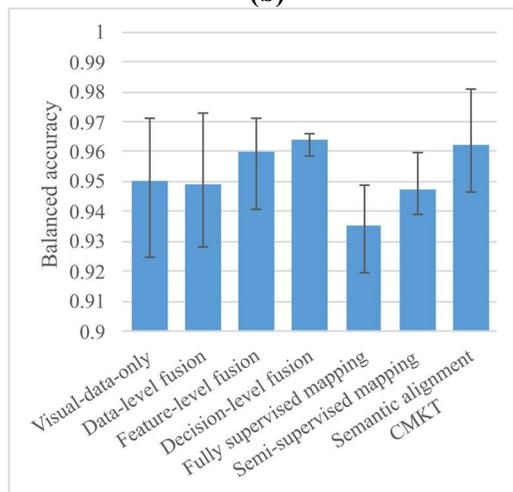

(c)

**Figure 8: Performance of the visual-data-only, multimodal fusion, and CMKT methods. (a) accuracy; (b) AUC-ROC; and (c) balanced accuracy. The CMKT models transferred knowledge from the audio to the visual modality.**

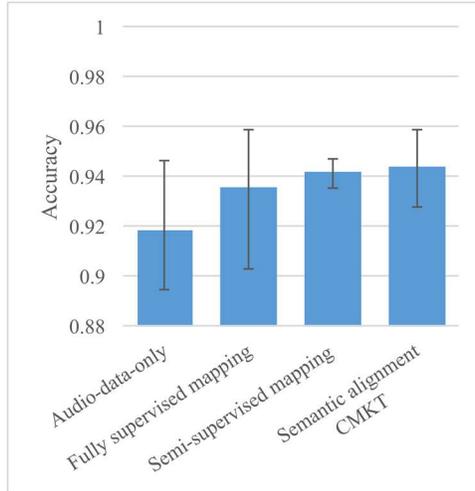

(a)

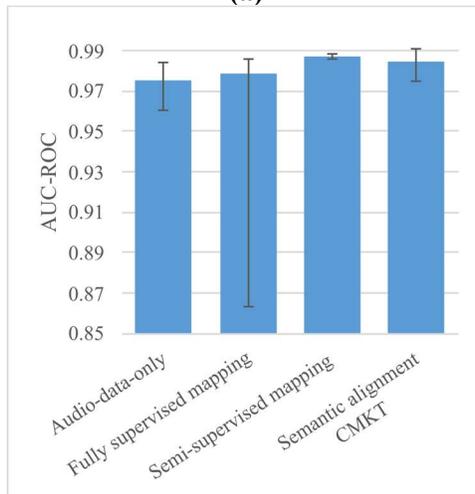

(b)

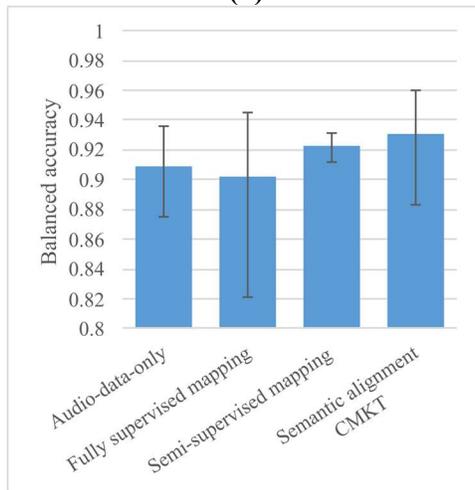

(c)

**Figure 9:** Performance of the audio-data-only and CMKT methods. (a) accuracy; (b) AUC-ROC; and (c) balanced accuracy. The CMKT models transferred knowledge from the visual to the audio modality.

## 5. Discussions

This section provides comprehensive discussions on the performance of the proposed CMKT methods regarding LAM in-situ defect detection. It first compares the overall predictive performance of the implemented methods. Then, the advantages of CMKT in ML-based process monitoring were investigated in detail, including enhanced denoising capability, effective feature extraction, higher separability, smaller data volume, shorter prediction runtime, and reduced costs. The advantages were illustrated using techniques such as XAI and dimensionality reduction. Finally, the limitations of CMKT were discussed.

### 5.1. Overall predictive performance

Figure 8 compares the visual-data-only, multimodal fusion, and CMKT methods. The visual-data-only single-modal method achieved an accuracy of 0.974, an AUC-ROC of 0.996, and a balanced accuracy of 0.971. Multimodal fusion provided higher optimal performance than the single-modal method, which aligned with the observations in [15]. The highest optimal accuracy (0.982) and AUC-ROC (0.999) among the multimodal fusion methods were obtained from feature-level fusion. The highest optimal balanced accuracy among the multimodal fusion methods was offered by data-level fusion (0.973). Decision-level fusion provided high average scores and small performance variations, which indicated that it was more consistent and robust. Overall, semantic alignment CMKT achieved the highest optimal accuracy, AUC-ROC, and balanced accuracy, which were 0.986, 0.9995, and 0.981. It also provided the highest or second-highest average and minimum accuracy, AUC-ROC, and balanced accuracy, which indicated its high robustness. The confusion matrix computed based on the test classification results with visual inputs (Figure A.1a) indicates that the model was not biased toward any visual group. Therefore, CMKT can provide comparable and even better predictive performance than multimodal fusion although it involves fewer modalities during the prediction phase. Nonetheless, the two mapping CMKT methods achieved lower scores than the visual-data-only method because they mapped the visual input to the audio modality. The hidden features extracted from the audio modality are usually less effective for LDED process defect detection than the visual modality [15].

Figure 9 compares the audio-data-only and CMKT methods. This figure does not include multimodal fusion because CMKT cannot achieve comparable predictive performance when solely taking audio data as input during the prediction phase. The audio-data-only single-modal method achieved an accuracy of 0.946, an AUC-ROC of 0.984, and a balanced accuracy of 0.936. Again, semantic alignment CMKT provided the highest optimal accuracy (0.958), AUC-ROC (0.991), and balanced accuracy (0.960). The confusion matrix computed based on the test classification results with audio inputs (Figure A.1b) indicates that the model was not biased toward any audio group. By mapping the audio data to the visual modality, both fully supervised and semi-supervised mapping CMKT achieved higher optimal and average accuracy and AUC-ROC than the audio-data-only method. However, their optimal balanced accuracies were similar to the audio-data-only method (around 0.94). Around 2% increase in accuracy and balanced accuracy was obtained by the semantic alignment CMKT method, which is equivalent to more than 25% error reduction compared to the audio-data-only method. Therefore, only the semantic alignment CMKT method showed a significant improvement in the predictive performance while transferring knowledge from the visual to the audio modality.

## 5.2. Computational complexity

Figure 10 and Figure 11 exhibit the training and prediction runtimes of the implemented methods. The training runtimes were the total runtimes accounting for all training epochs, while the prediction runtimes accounted for one forward pass of the validation set through the trained models. The ML models were trained on a Windows 11 system with an Intel Core i7 12$^{th}$ Gen processor and an NVIDIA GeForce RTX 4090 GPU, utilizing Python 3.6 and the PyTorch framework. CUDA was employed for GPU acceleration to enhance computational performance during training. The two single-modal methods had the shortest training runtimes with average training runtimes below 70 seconds because they only involved one modality. The average training runtime of data-level fusion (around 400 seconds) was three times higher than feature-level and decision-level fusion. More channels were deployed in the convolutional modules of the data-level fusion models where the two modalities were concatenated at the input layer. The other multimodal fusion methods concatenated the two modalities after the convolutional modules, which reduced the number of weights. The average training runtime of fully supervised mapping (around 80 seconds) was close to feature-level and decision-level fusion (around 100 seconds). The average training runtime of semi-supervised mapping was twice that of fully supervised mapping because semi-supervised mapping trained two autoencoders while fully supervised mapping only trained one CNN in Phase 1. The average training runtime of semantic alignment CMKT was the highest among all methods, which was around 750 seconds for both knowledge transfer directions. Semantic alignment CMKT had different training and prediction runtimes in Figure 10 and Figure 11 because two batches of the top 50 models were separately chosen using the visual and audio validation sets. Semantic alignment CMKT employed semantic alignment loss and separation loss that calculated pairwise distances in the encoded space. Besides, the convolutional encoder and task classifier were updated separately using different loss terms, leading to more computations per epoch and a higher number of epochs to converge.

Although semantic alignment CMKT yielded high training runtimes, it had short prediction runtimes (around 0.15 seconds), which were close to the single-modal methods. A semantic alignment CMKT model was equivalent to a single-modal CNN model during the prediction phase because no semantic alignment loss or separation loss was computed. The average prediction runtimes of multimodal fusion methods were one to three times higher than semantic alignment CMKT because they utilized both modalities when making predictions, while semantic alignment CMKT only involved one modality. The cross-modality mapping methods also had longer prediction runtimes than semantic alignment because they employed multiple networks.

## 5.3. Investigation based on explainable artificial intelligence

To investigate how semantic alignment CMKT enhances model performance, we implemented LIME, an XAI method, that visualizes features learned by a trained model [86]. Mathematically, LIME seeks to find an interpretable surrogate model that can locally approximate the behavior of the complex model around a specific input $x_0$. Let $f$ represent the original, complex model, and $g$ represent the simpler, interpretable surrogate model (e.g., a linear or decision tree model) that LIME uses to approximate $f$ locally. The objective of LIME is to minimize the following loss function:

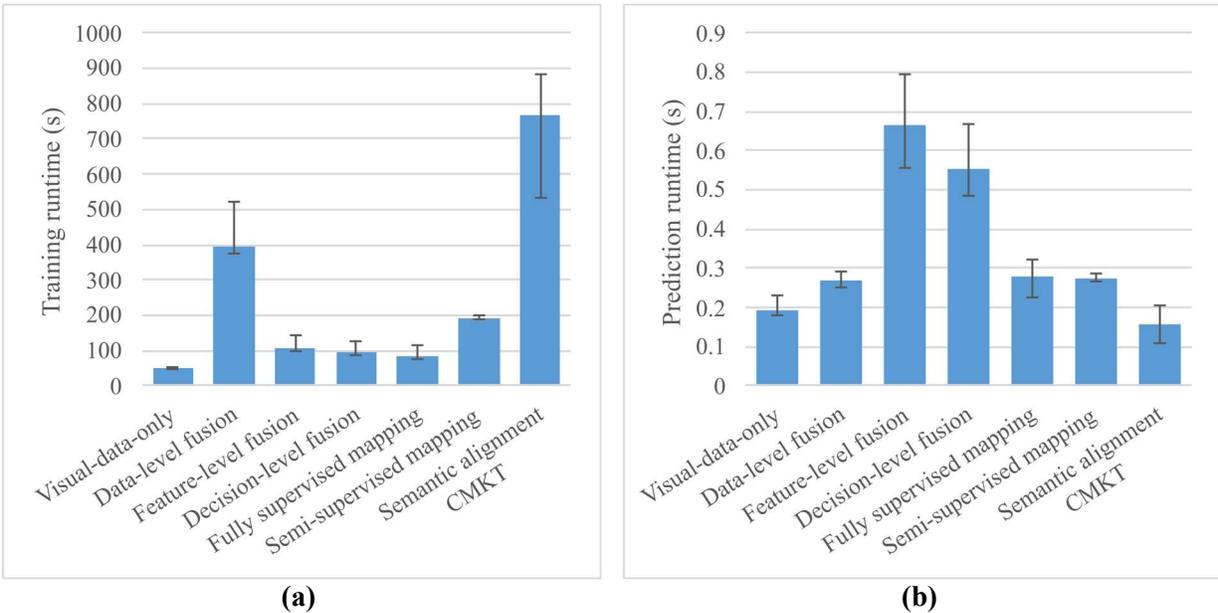

(a)  (b)

**Figure 10: Runtimes of the visual-data-only, multimodal fusion, and CMKT methods. (a) training runtime and (b) prediction runtime. The CMKT models transferred knowledge from the audio to the visual modality.**

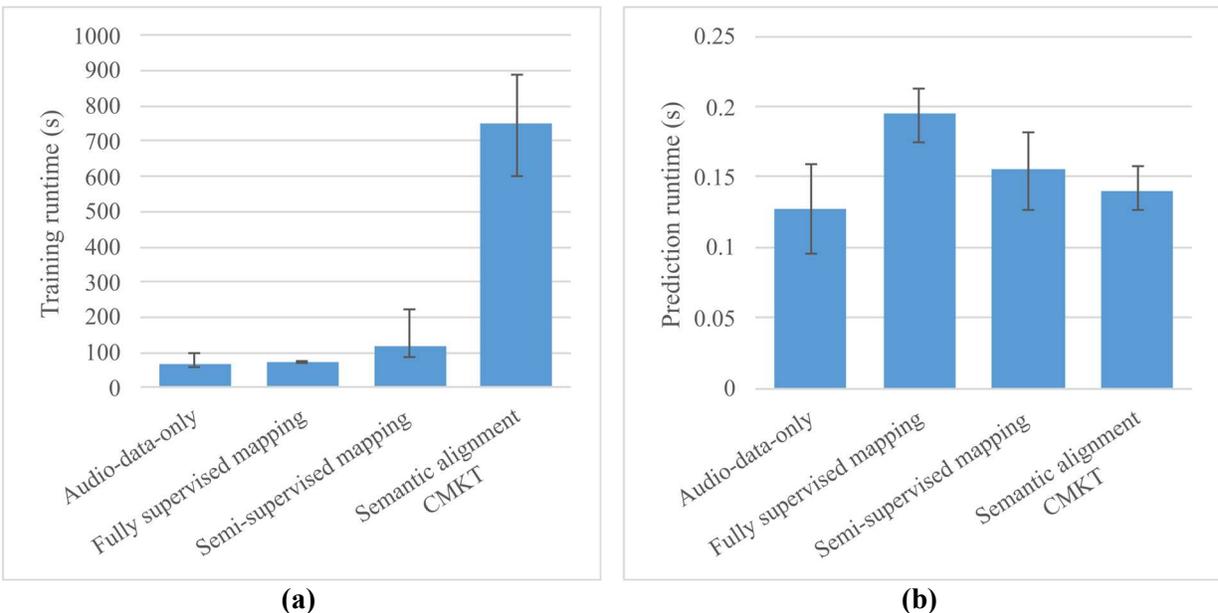

(a)  (b)

**Figure 11: Runtimes of the audio-data-only and CMKT methods. (a) training runtime and (b) prediction runtime. The CMKT models transferred knowledge from the visual to the audio modality.**

$$L(f, g, \pi_x) + \Omega(g), \tag{12}$$

where $L(f, g, \pi_x)$ is a local loss function that measures how closely the predictions of the interpretable model $g$ match those of the complex model $f$ in the local neighborhood of the input $x_0$; $\pi_x$ is a proximity measure that assigns more weight to points closer to $x_0$ in the input space, ensuring that the surrogate model $g$ is more accurate near $x_0$; and $\Omega(g)$ is a complexity penalty on the interpretable model $g$, encouraging it to remain simple and interpretable.

For image inputs like melt pool images and spectrograms, LIME applies the above formulation by first segmenting an image input $x_0$ into superpixels. It then perturbs the image by randomly altering these superpixels and observes how these changes affect the predictions of the complex model $f$. Let $x_i'$ represent the perturbed input image after altering the $i^{th}$ superpixel. The local surrogate model $g$ learns the contribution of each superpixel by evaluating the changes in predictions of $f(x_i')$. LIME then fits the surrogate model $g$ based on these perturbations and generates a feature importance map that highlights the most influential superpixels for the model prediction at $x_0$. The superpixels that contribute the most to the prediction are assigned higher weights, and these are displayed in the form of an explanation. By capturing the local behavior of the model around a specific image and evaluating the importance of different superpixels, LIME offers an interpretable explanation for individual predictions. This helps validate whether the model is focusing on meaningful parts of the image or whether it is influenced by irrelevant features [87].

In this study, we compared visual-data-only and audio-data-only models with semantic alignment CMKT models using LIME for explainability. A ridge regression model was selected as the simple surrogate model because it inherently provides both the local loss function and the complexity penalty. Cosine distance was selected to compute the proximity between $x_0$ and $x_i'$ for weight assignment due to its robustness against lighting condition variations [88]. Once the LIME result was computed for an image, the top five superpixels that positively contributed to the model prediction were highlighted, forming the positive mask ($P_{lime}$) (Figure 12). Since these highlighted superpixels represent the most influential features that led to the model prediction, they can be considered the most representative features the model learned to extract from the input data.

Figure 12a presents examples of the positive masks of visual inputs generated based on the visual-data-only and semantic alignment CMKT models. It can be observed that both models primarily extracted the contours of the melt pools rather than the entire melt pools. This is because the melt pools exhibit excessive luminance, exceeding the dynamic range of the camera, which leads to image saturation. As a result, pixel values in these saturated areas reach their maximum with minimal variance, providing little useful information. Therefore, the contours of the melt pools become the most representative features. The key difference between the two models was that the visual-data-only model always extracted the rim enclosing the melt pools, while the semantic alignment CMKT model learned to neglect most of them. The presence of the rim is caused by reflections of the melt pool on the LDED nozzle due to the coaxial configuration of the CCD camera. These rims introduce significant noise into the melt pool images and are considered task-irrelevant or redundant features. The denoising capability of semantic alignment CMKT will be discussed in detail in Section 5.4.1.

Figure 12b presents examples of the positive masks of audio inputs generated based on the audio-data-only and semantic alignment CMKT models. In contrast to melt pool images, it is challenging to visually interpret and understand the extracted features from the spectrograms. Therefore, a frequency analysis was conducted in Section 5.4.2, revealing the capability of the semantic alignment CMKT model to extract salient audio features.

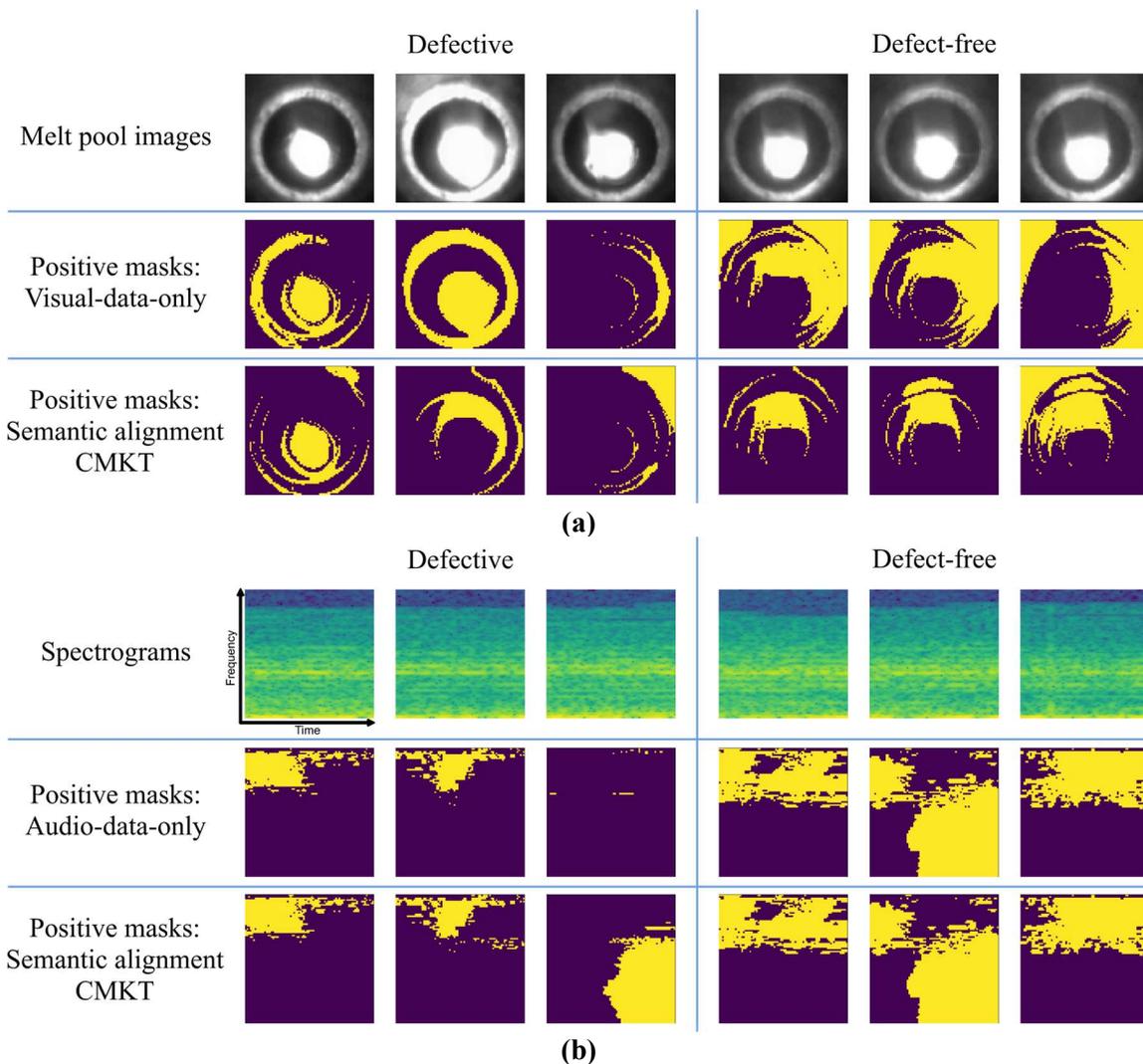

Figure 12: Examples of positive masks ($P_{lime}$) in yellow generated using LIME for a) the visual modality (melt pool images) and b) the audio modality (spectrograms).

## 5.4. Advantages of cross-modality knowledge transfer

The advantages of the proposed CMKT methods are discussed with respect to their enhanced denoising capability, effective extraction of salient features, improved separability, and higher operational efficiency.

### 5.4.1. Noise reduction

In multimodal monitoring setups, researchers deploy multiple sensors to capture certain physical phenomena that contain features relevant to the modeling task, while often overlooking the inevitable noise introduced by each modality [89]. In the visual modality, optical imaging issues such as random noise and flares can contaminate the morphological features by blurring the contours of melt pools and plumes [90]. In the audio modality, random and background noises interfere with the acoustic signatures of the LDED process [91]. These types of noises increase the difficulty of extracting useful features and consume the learning capacity of the ML model, ultimately reducing prediction performance. Moreover, it is often challenging to completely segregate and remove these noises. In this study, the reflection on the nozzle was subject to heavy noise and contributed little useful information to the modeling task. We generated a nozzle mask ($P_{nozzle}$) to indicate the location of the reflection, as its position was relatively stable. However, removing the reflection from the images could be challenging because it sometimes overlaps with the melt pool. Simply cutting out $P_{nozzle}$ from the images could distort the natural morphological characteristics of the melt pools, making it an impractical solution.

Semantic alignment CMKT can extract denoised representations by leveraging the similarities between two modalities. Section 3.2.1 shows that semantic alignment CMKT employs contrastive and semantic alignment loss that forces the model to extract modal-shared and task-relevant features. This method implicitly filters out noises, which are often modal-specific and task-irrelevant. The denoising capability of semantic alignment CMKT was validated using the pipeline illustrated in Figure 13. To examine the visual-data-only model, a positive mask ($P_{lime}$) was generated for each melt pool image from the visual test set ($D_{V,test}$) using LIME to highlight the features extracted by the model. Thereafter, the intersection of the nozzle and LIME positive mask ($P_{nozzle} \cap P_{lime}$) was computed and quantified by counting the number of highlighted pixels. In this way, the larger the intersection, the more the model learned to extract noises from the reflection on the nozzle. The same pipeline was applied to compute the intersections for the semantic alignment CMKT model, providing a comparison of their capabilities to avoid learning noise-related features.

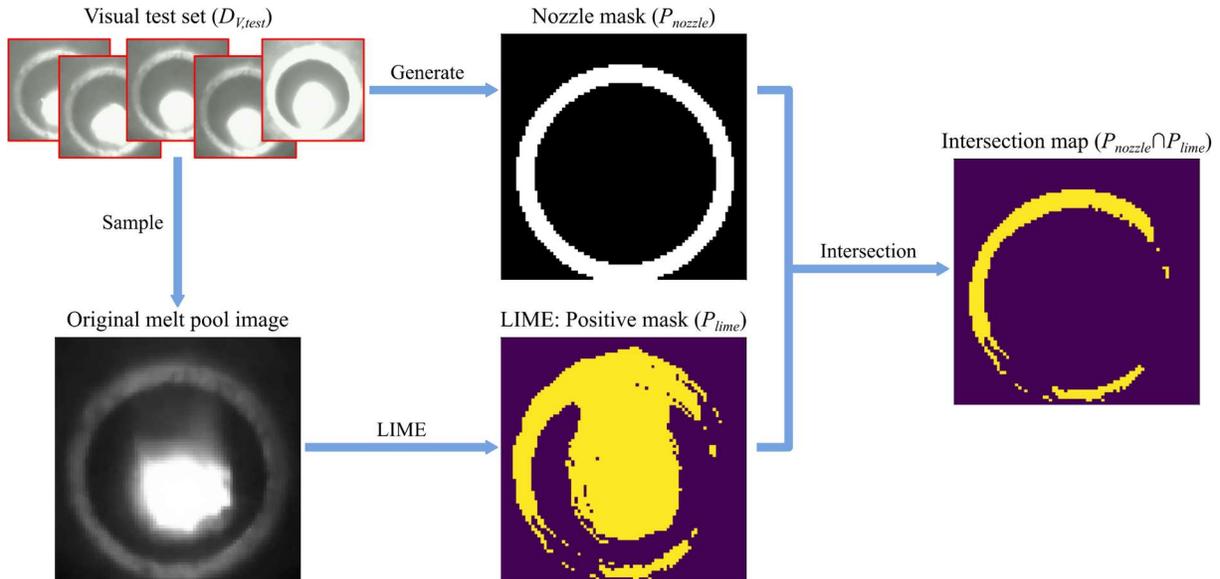

**Figure 13: Generation of the intersection maps of the test set to compare the denoising capabilities of the visual-data-only and semantic alignment CMKT models.**

The distributions of the intersections between the nozzle mask and the LIME positive mask for both models were compared in Figure 14. Probability density functions at every 150$^{th}$ epoch were calculated using kernel density estimation. Distributions closer to the left (i.e., closer to zero) indicate that fewer noises from the nozzle reflection were extracted. At the 150$^{th}$ epoch, both models had an average of over 500 pixels, with the visual-data-only model showing less variance. As training continued, the distribution for semantic alignment CMKT rapidly shifted leftward, suggesting an improvement in denoising capability, whereas the visual-data-only distribution slightly shifted leftward. At the 1200$^{th}$ epoch, the average number of pixels of the semantic alignment CMKT model decreased to 358 while the average of the visual-data-only model was 547. Additionally, we manually removed the reflection on the nozzle by cutting $P_{nozzle}$ out from each melt pool image. Based on this modified dataset, we conducted a hyperparameter search to train the optimal visual-data-only model, which achieved a test accuracy of 0.981. This test accuracy was higher than the original visual-data-only model (0.974) but still lower than the semantic alignment CMKT model (0.986). This highlights that semantic alignment CMKT naturally focuses on modal-shared, task-relevant features and disregards noise, outperforming manual noise removal.

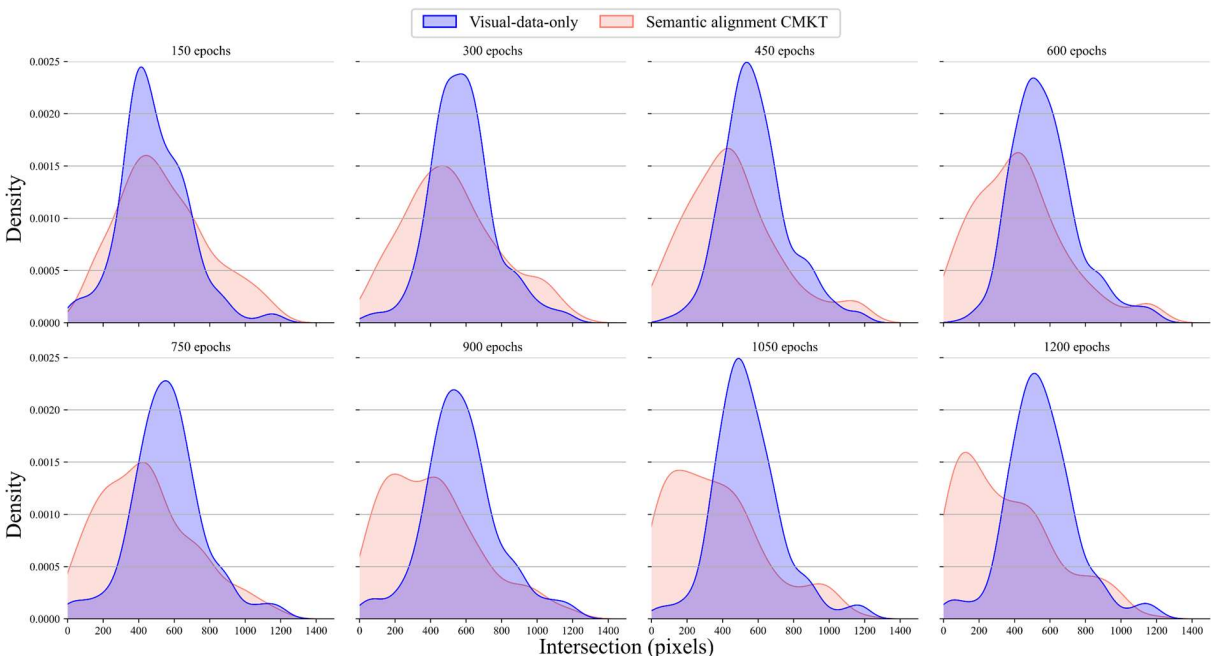

**Figure 14: Estimated probability density functions of the number of pixels included in the intersection maps for the visual-data-only and semantic alignment CMKT models.**

The noise reduction capability of semantic alignment CMKT was further validated by comparing the implemented methods with respect to varied visual and audio noise levels (Figure 15). To generate visual datasets of different noise levels, additive white Gaussian noise (AWGN) defined by its mean and standard deviation was applied to each example of the original visual dataset. Five synthetic visual datasets were generated based on five Gaussian distributions with the same mean of zero and five different standard deviations {5, 10, 15, 20, 25}. A large standard deviation indicates a larger noise intensity, measured by the pixel intensity of grayscale images, which ranges from 0 to 255. AWGN was also utilized to generate five synthetic audio datasets with the same mean of zero and five different standard deviations, dictated by five signal-to-noise ratios {70, 65, 60, 55, 50} measured in dB. Given a fixed signal power, the larger the signal-to-noise ratio, the smaller the noise. The five signal-to-noise ratios representing negligible to small noise levels were selected based on the high-quality microphone.

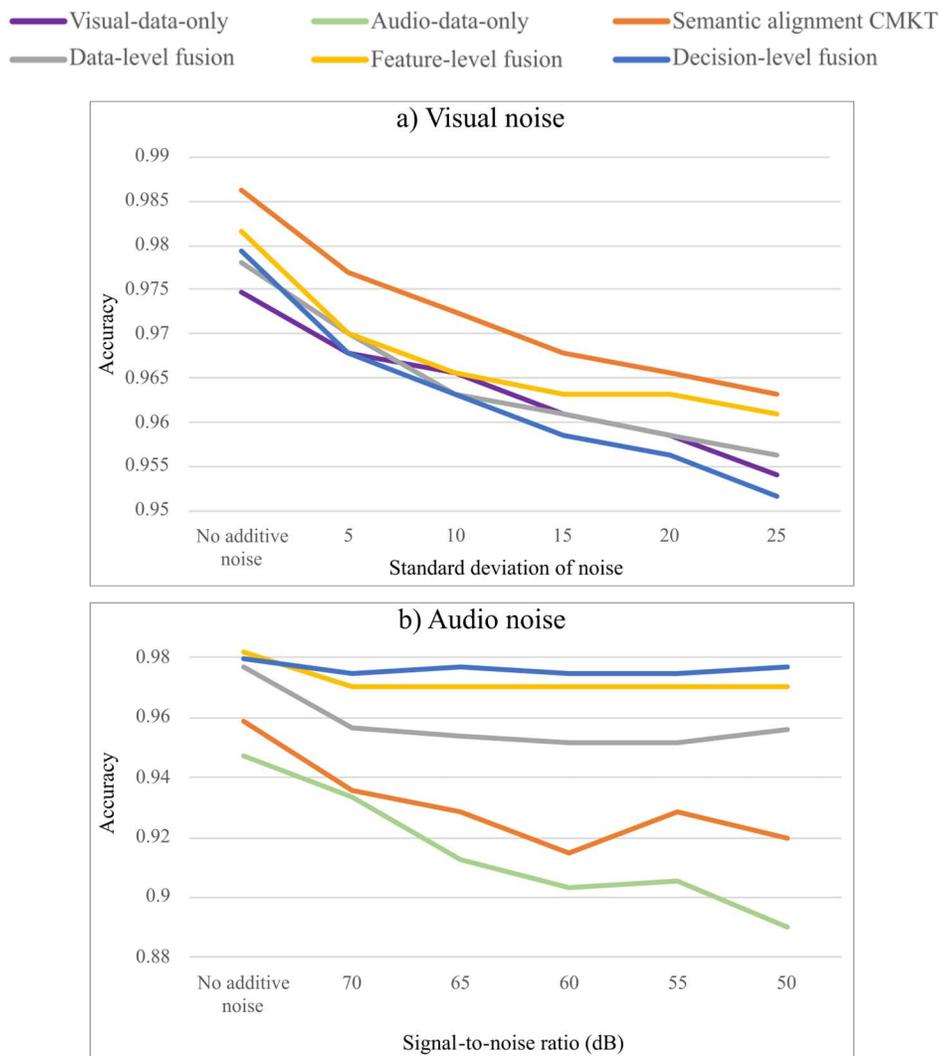

**Figure 15: Performance shift of the implemented methods caused by additive a) visual noise and b) audio noise.**

Models of different methods were trained using the optimal hyperparameters found in Section 4 for each noise level. Figure 15a presents the accuracy of each method as the visual noise level increases with no additive noise applied to the audio modality. Semantic alignment CMKT obtained the highest accuracies across all noise levels investigated, indicating its strong noise reduction capability. Feature-level fusion performed better than the visual-data-only and other multimodal fusion methods because the multimodal features it extracts can potentially filter out noises. Although data-level fusion can also extract multimodal features, its denoising capability is limited due to the high input dimensionality. Figure 15b exhibits the accuracy of each method as the audio noise level increases with no additive noise applied to the visual modality. Semantic alignment CMKT performed better than the audio-data-only method, especially for signal-to-noise ratios from 65 to 50. When only feeding audio data into semantic alignment CMKT models, they cannot achieve the same accuracy as multimodal fusion methods, as indicated in Section 5.1. The feature-level fusion and decision-level fusion can maintain high accuracy because they have access to the information-rich visual modality with no additive noise. Again, the compromised performance of the data-level fusion method might be attributed to the high input dimensionality. In conclusion, this analysis further validates the noise reduction capability of semantic alignment CMKT.

### 5.4.2. Extraction of salient features

The capability of semantic alignment CMKT to effectively extract salient features was illustrated using LIME and frequency analysis (Figure 16). For the audio-data-only model, an LIME positive mask was generated for each spectrogram in the audio test set. We equally segregate the entire bandwidth (0 to 22000 Hz) into 80 frequency ranges, each covering 275 Hz. For each frequency range in a positive mask, we identified whether any pixel was highlighted. If so, the corresponding frequency range was marked as highlighted. This reduced the matrix of positive masks to a two-dimensional shape of 80 × 435 (the test sample size). Thereafter, we counted the number of samples that had highlighted pixels regarding each frequency range. Finally, the distributions of the highlighted frequency ranges can be visualized using histograms (Figure 17). The same process was repeated to plot the histograms of the highlighted frequency ranges for the semantic alignment CMKT model.

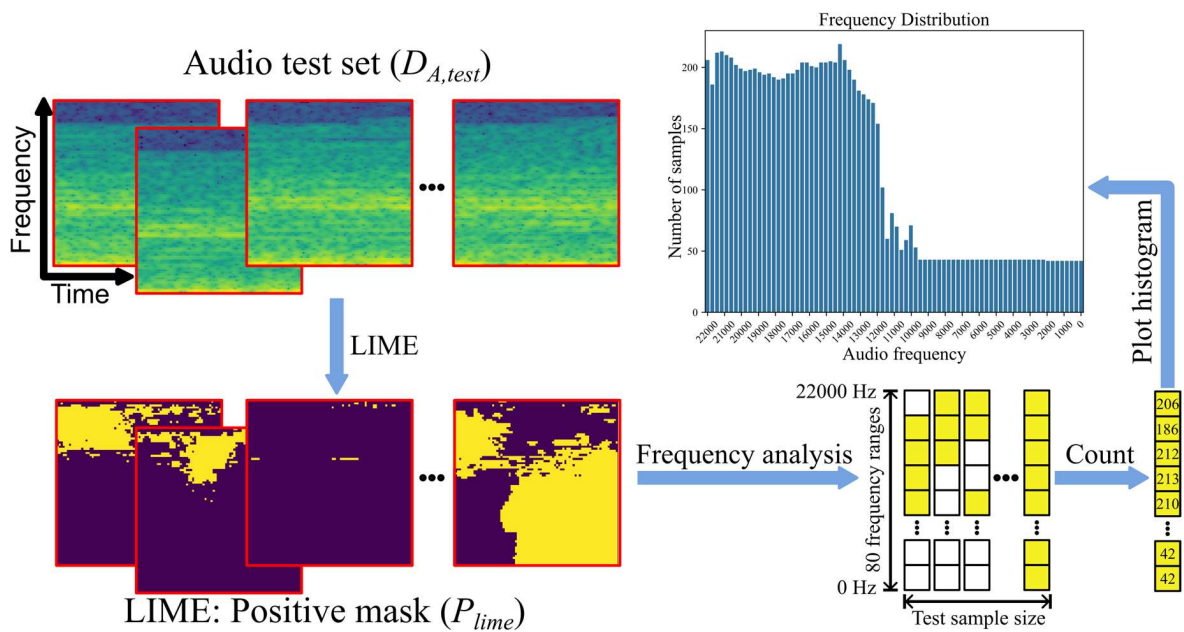

**Figure 16: Frequency analysis pipeline to investigate features extracted from the audio modality.**

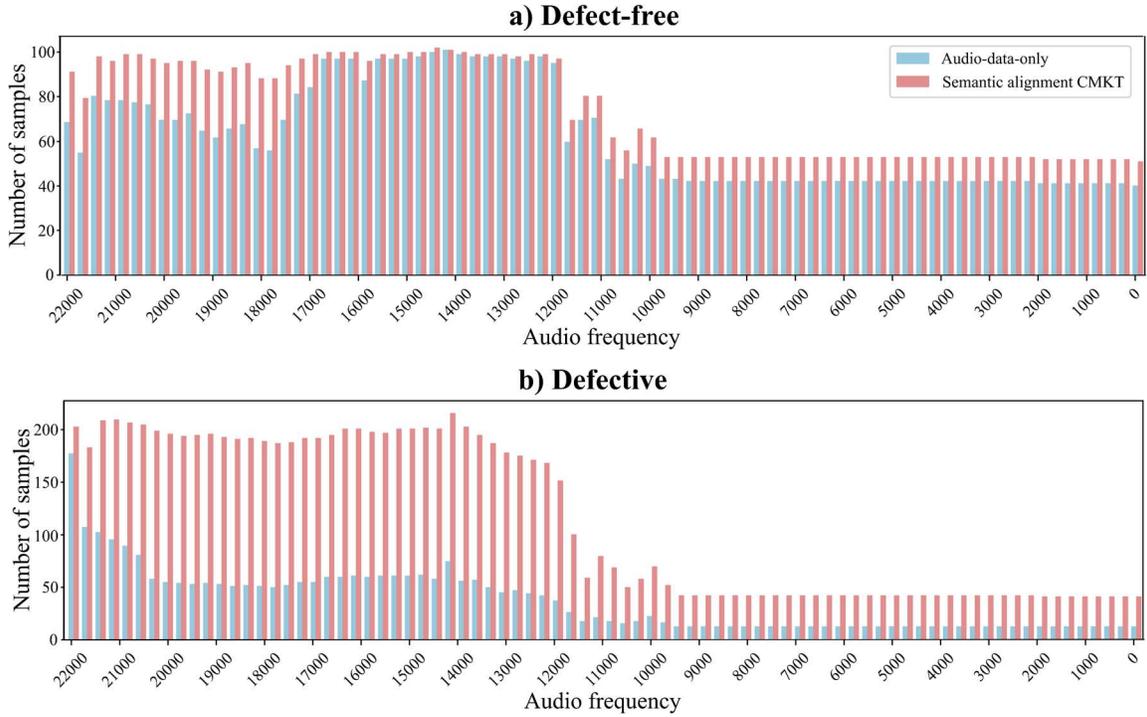

Figure 17: Histograms of the highlighted frequency ranges in the LIME positive masks: a) defect-free class and b) defective class.

Both the low-frequency (0 to 11000 Hz) and high-frequency (11000 Hz to 22000 Hz) ranges were found to contain essential information for defect detection, as features were extracted across the entire bandwidth. Notably, more features were drawn from the high-frequency range than from the low-frequency range. This could be attributed to the resolution of the spectrograms, where each bin representing 275 Hz may have failed to capture subtle pitch changes that are critical in the low-frequency ranges. Compared with the audio-data-only model, the semantic alignment CMKT model extracted marginally more features from the defect-free examples (Figure 17a) and significantly more features from the defective examples (Figure 17b). An investigation revealed that both models typically relied on fewer than five superpixels to support the defective class predictions. However, the semantic alignment CMKT model often captured more superpixels than the audio-data-only model, leading to fewer false negatives by forming a more complete understanding of the defective class. In conclusion, semantic alignment CMKT can extract more salient representations by leveraging features shared across modalities, offering improved performance in feature extraction for defect detection.

### 5.4.3. Improved separability

In Figure 18a, the encoded representations of the test data in the semantic alignment CMKT model were reduced to two dimensions using t-SNE. The cluster contains four groups: the audio defect-free, audio defective, visual defect-free, and visual defective groups. Compared with the audio modality, the visual defect-free and defective classes were well split with marginal overlap. This was also reflected in Figure 8 and Figure 9 where the semantic alignment models achieved higher performance when evaluated on visual data than audio data. It was observed that semantic alignment CMKT encouraged the visual defect-free and defective groups to split away from each other because of the audio data. The audio data became a buffer zone that increased the separability between the visual defect-free and defective groups. The improved separability facilitates the classification between different visual groups and increases model robustness against noise and outliers. Figure 18b visualizes the encoded representations of the test data in the last hidden layer of the optimal feature-level fusion model using t-SNE. Compared with semantic alignment CMKT, the fused defect-free and defective classes were close to each other with a long borderline. Therefore, feature-level fusion models were more prone to outliers than semantic alignment CMKT models.

The cross-modality mapping CMKT methods offered performance improvement when transferring knowledge from the visual modality to the audio modality. This occurred because the visual features were more effective for LDED in-situ defect detection than the audio features. Therefore, the predictive performance was improved when the visual features were derived from the audio modality using cross-modality mapping. However, the performance increase obtained from cross-modality mapping was not as high as that of semantic alignment CMKT.

### 5.4.4. Improved operational efficiency

CMKT can reduce the data volume and computational complexity of ML-based AM process monitoring systems. After transferring knowledge from the source to the target modality, the source modality can be removed during the prediction phase (i.e., operation). This decreases data volume and execution time for data acquisition, transmission, and preprocessing before making predictions. Thus, the system demands a less powerful computer for local computing and allows faster data transmission to the cloud in an IoT environment. Besides, the architecture of the CMKT models can be more lightweight than multimodal fusion models due to fewer modalities, leading to shorter prediction runtimes (Figure 10 and Figure 11). Additionally, CMKT decreases the hardware, operation, and maintenance costs by reducing the number and types of sensors, thus enhancing the scalability of LAM. For instance, the monitoring system utilized in this research, as described in Section 3.1, employs a CCD camera subsystem and a microphone subsystem, each costing around $5000. Due to the coaxial setup of the CCD camera, removing the microphone from the monitoring system after CMKT would be a more feasible option. This microphone can be reused for other LAM machines in the workshop to improve their process monitoring performance using CMKT without extra sensor costs.

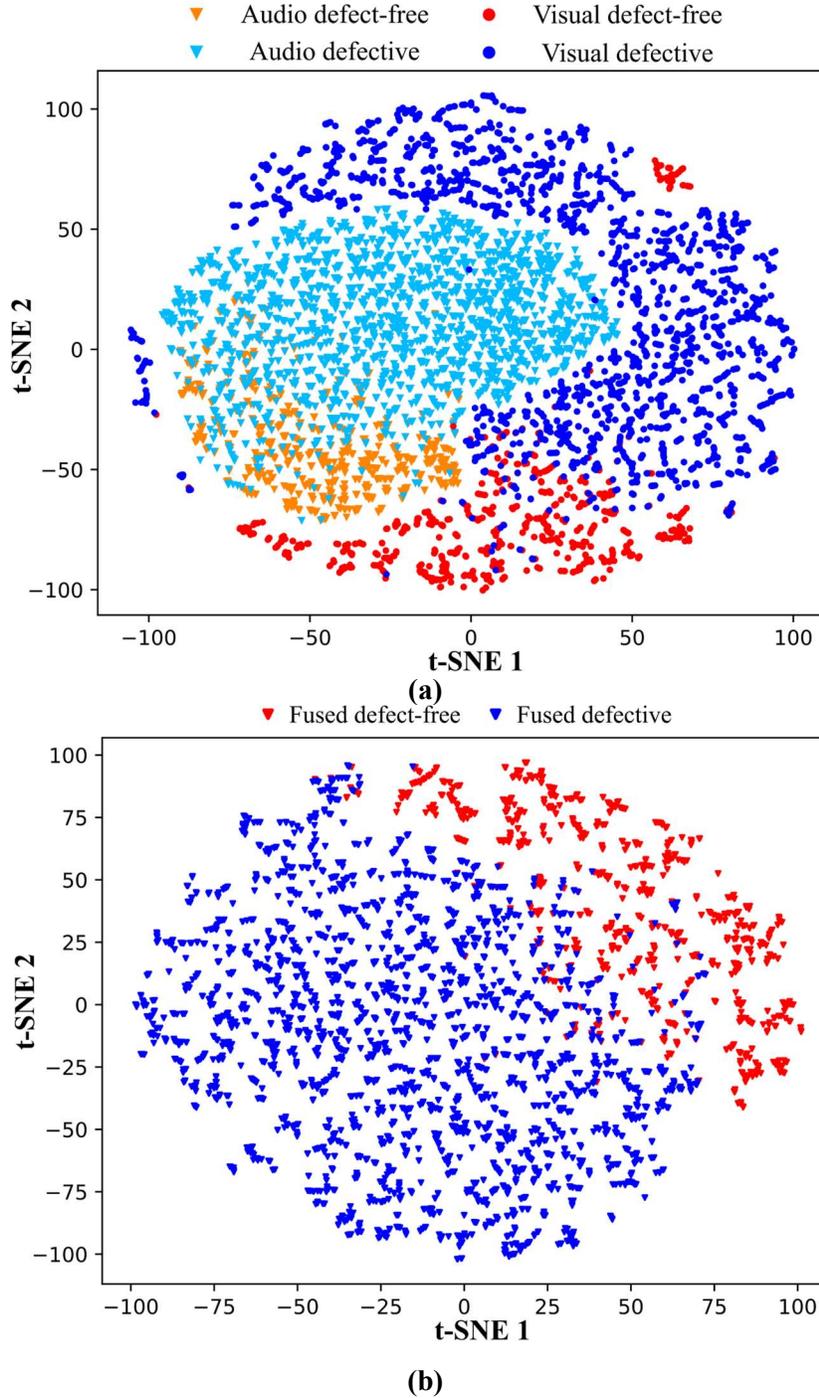

**Figure 18: T-SNE visualization of the encoded space of (a) semantic alignment CMKT and (b) feature-level multimodal fusion.**

### 5.5. Limitations of cross-modality knowledge transfer

The key underlying assumption of CMKT is that the source and target modalities contain overlapping information about the prediction task. Visual features can be extracted from the visual modality and audio features can be extracted from the audio modality. By transferring knowledge from the source to the target modality, the model can extract more useful features from the target modality than single-modal methods. However, CMKT will not provide performance improvements if the two modalities contain only mutually exclusive information (i.e., none of the features extracted from one modality are derivable from the other modality) [92]. Further, the training process of CMKT methods is usually more complicated than multimodal fusion methods. When deploying CMKT for industrial LAM monitoring systems, it is crucial to involve ML experts with strong knowledge transfer expertise to mitigate the risk of negative transfer, which could compromise the performance of ML models. For instance, cross-modality mapping methods might lower the accuracy of defect detection in this study compared with single-modal methods, whereas semantic alignment CMKT instead improved the performance. This requires systematic implementation and comparison of multiple CMKT methods to obtain optimal performance improvements.

## 6. Conclusions

This paper introduced CMKT methods that transfer knowledge from the source to the target modality during the training phase, therefore removing the source modality during the prediction phase. In an LDED in-situ defect detection case study, the proposed CMKT methods, including semantic alignment, fully supervised mapping, and semi-supervised mapping, were compared with single-modal (i.e., visual-data-only and audio-data-only) and multimodal fusion (i.e., data-level, feature-level, and decision-level fusion) methods. The highest optimal accuracy (98.6%), AUC-ROC (0.9995), and balanced accuracy (98.1%) were obtained using the semantic alignment CMKT method. While training the semantic alignment CMKT models required more time, the prediction runtime was considerably shorter than that of the multimodal fusion models and similar to that of the single-modal methods. Using LIME, we visualized the features extracted by the single-modal and semantic alignment CMKT models from both the visual and audio modalities. We discovered that semantic alignment CMKT exhibits robust denoising capabilities. Specifically, this method was able to disregard irrelevant reflections in melt pool images caused by the nozzle. It also demonstrated exceptional feature extraction capabilities, particularly in identifying features that provided a more comprehensive understanding of the defective class, reducing false negatives. Additionally, semantic alignment CMKT offered better separability and significantly reduced the computational complexity and costs during operation. The limitation of CMKT is that the two modalities must contain overlapping information about the prediction task. Otherwise, knowledge may not be transferable between the source and the target modalities. Future works will extend the proposed CMKT methods to other LAM processes such as LPBF and WAAM. Other modalities such as thermal imaging and process parameters will also be investigated for CMKT.

**Author contributions: CRediT**

**Jiarui Xie:** Conceptualization, Methodology, Software, Validation, Formal analysis, Investigation, Writing - Original Draft, Visualization.

**Mutahar Safdar:** Conceptualization, Methodology, Software, Validation, Writing - Original Draft.

**Lequn Chen:** Conceptualization, Investigation, Resources, Data Curation, Writing - Original Draft, Supervision, Project administration.

**Seung Ki Moon:** Conceptualization, Resources, Writing - Review & Editing, Supervision.

**Yaoyao Fiona Zhao:** Conceptualization, Resources, Writing - Review & Editing, Supervision, Funding acquisition.

**Funding sources**

This work is funded by McGill University Graduate Excellence Fellowship Award [grant number 00157]; Mitacs Accelerate program [grant number IT13369]; McGill Engineering Doctoral Award (MEDA); and National Research Council Canada [NRC INT-015-1]. It is also supported by Agency for Science, Technology and Research (A*STAR) of Singapore through RIE2025 MTC IAF-PP grant (Grant No. M22K5a0045).

**Data Availability**

The dataset used in this work is open-source and has been publicly released at Zenodo: https://zenodo.org/records/12604782.

**Appendix**

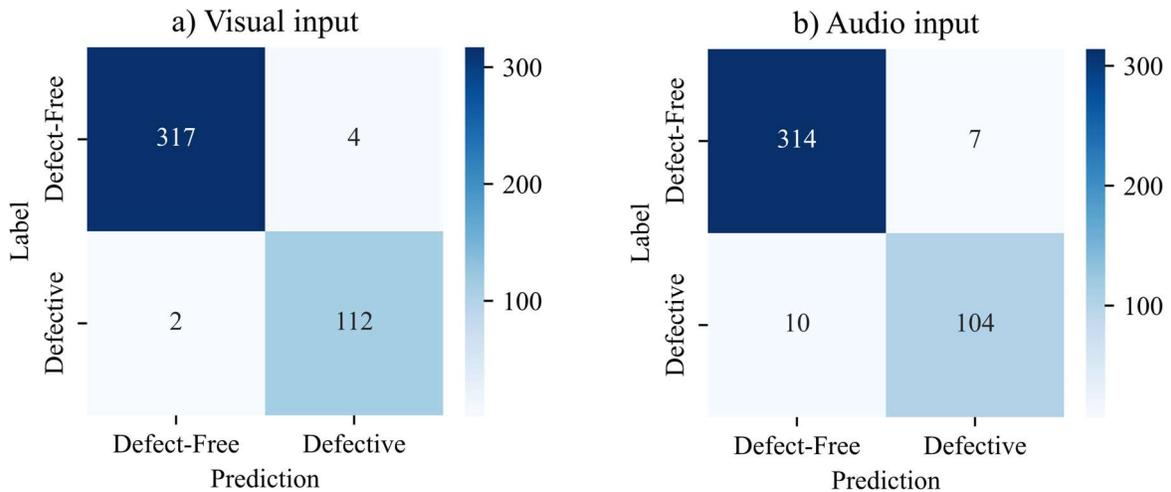

**Figure A.1: Confusion matrices based on the test results of the semantic alignment CMKT model with a) Visual input and b) Audio input.**

**Table A.1: The optimal hyperparameters of the fully supervised mapping method. The models transferred knowledge from the visual to the audio modality.**

| | |
|---|---|
| **Common settings** | |
| Optimizer | Adam |
| Class weighting | defective/defect-free=1/3 |
| **Phase 1: Feature learning ($G_V$)** | |
| Initial learning rate | 0.0012911986 |
| Loss | Binary cross entropy |
| Input and output layer dimensions | 80 × 80 and 1 |
| Number of epochs | 500 |
| *Layer* | *Hyperparameters* |
| 1st Convolutional module | Conv2d (30 channels, kernel size=4, stride=1, padding= 'same'), ReLU(), MaxPool2d (kernel size=2, stride=1) |
| 2nd Convolutional module | Conv2d (30 channels, kernel size=4, stride=1, padding= 'same'), ReLU(), MaxPool2d (kernel size=1, stride=1) |
| 3rd Convolutional module | Conv2d (30 channels, kernel size=4, stride=1, padding= 'same'), ReLU(), MaxPool2d (kernel size=2, stride=1) |
| 4th Convolutional module | Conv2d (57 channels, kernel size=4, stride=1, padding= 'same'), ReLU(), MaxPool2d (kernel size=2, stride=1) |
| Flatten | Flatten (57 × 10 × 10) |
| 1st Fully connected module | Linear (57 × 10 × 10, 380), ReLU(), Dropout (0.02474) |
| 2nd Fully connected module | Linear (380, 164), ReLU(), Dropout (0.02474) |
| 3rd Fully connected module | Linear (164, 134), ReLU(), Dropout (0.02474), BatchNorm1d |
| 4th Fully connected module | Linear (134, 1), sigmoid |
| **Phase 2: Mapping ($G_{A1}$)** | |
| Initial learning rate | 0.0026540529 |
| Loss | Mean squared error |
| Input and output layer dimensions | 80 × 80 and 15 |
| Number of epochs | 1500 |
| *Layer* | *Hyperparameters* |
| 1st Convolutional module | Conv2d (45 channels, kernel size=2, stride=1, padding= 'same'), ReLU(), MaxPool2d (kernel size=2, stride=1) |
| 2nd Convolutional module | Conv2d (45 channels, kernel size=2, stride=1, padding= 'same'), ReLU(), MaxPool2d (kernel size=1, stride=1) |
| 3rd Convolutional module | Conv2d (45 channels, kernel size=2, stride=1, padding= 'same'), ReLU(), MaxPool2d (kernel size=2, stride=1) |
| 4th Convolutional module | Conv2d (54 channels, kernel size=2, stride=1, padding= 'same'), ReLU(), MaxPool2d (kernel size=2, stride=1) |
| Flatten | Flatten (54 × 10 × 10) |
| 1st Fully connected module | Linear (64 × 10 × 10, 418), ReLU(), Dropout (0.02955) |
| 2nd Fully connected module | Linear (418, 128), ReLU(), Dropout (0.02955) |
| 3rd Fully connected module | Linear (128, 147), ReLU(), Dropout (0.02955) |
| 4th Fully connected module | Linear (147, 132), ReLU(), Dropout (0.02955) |
| 5th Fully connected module | Linear (132, 78), ReLU(), Dropout (0.02955) |
| 6th Fully connected module | Linear (78, 15) |

| Phase 3: Finetuning ($G_{A2}$) | |
|---|---|
| Initial learning rate | 0.0059795128 |
| Loss | Binary cross entropy |
| Input and output layer dimensions | 15 and 1 |
| Number of epochs | 150 |
| *Layer* | *Hyperparameters* |
| 1st Fully connected module | Linear (15, 144), ReLU(), Dropout (0.04803) |
| 2nd Fully connected module | Linear (144, 108), ReLU(), Dropout (0.04803) |
| 3rd Fully connected module | Linear (108, 97), ReLU(), Dropout (0.04803) |
| 4th Fully connected module | Linear (97, 36), ReLU(), Dropout (0.04803) |
| 5th Fully connected module | Linear (36, 102), ReLU(), Dropout (0.04803) |
| 6th Fully connected module | Linear (102, 1), sigmoid() |

**Table A.2: The optimal hyperparameters of the semi-supervised mapping method. The models transferred knowledge from the visual to the audio modality.**

| Common settings | |
|---|---|
| Optimizer | Adam |
| Class weighting | defective/defect-free=1/3 |
| Phase 1: Feature learning ($G_V$) | |
| Initial learning rate | 0.00105828121305257 |
| Loss | Mean Squared Error |
| Input and output dimensions | 80 × 80 and 80 × 80 |
| Number of epochs | 500 |
| *Layer* | *Hyperparameters* |
| Encoder Module: 1st Convolutional layer | Conv2d (21 channels, kernel size=3, stride=2, padding=1), ReLU () |
| Encoder Module: 2nd Convolutional layer | Conv2d (128 channels, kernel size=3, stride=2, padding=1), ReLU () |
| Encoder Module: 3rd Convolutional layer | Conv2d (100 channels, kernel size=3, stride=2, padding=1), ReLU () |
| Flatten | Flatten (100 × 10 × 10) |
| Encoder Fully Connected Layer | Linear (10000, 42), ReLU() |
| Decoder Fully Connected Layer | Linear (42, 10000), ReLU() |
| Unflatten | Unflatten (100 × 10 × 10) |
| Decoder Module: 1st Deconvolutional layer | ConvTranspose2d (105 channels, kernel size=3, stride=2, padding=1, output padding=1), ReLU() |
| Decoder Module: 2nd Deconvolutional layer | ConvTranspose2d (56 channels, kernel size=3, stride=2, padding=1, output padding=1), ReLU() |
| Decoder Module: 3rd Deconvolutional layer | ConvTranspose2d (1 channels, kernel size=3, stride=2, padding=1, output padding=1), Sigmoid() |

| Phase 1: Feature learning ($G_A$) | |
|---|---|
| Initial learning rate | 0.000926110702226012 |
| Loss | Mean Squared Error |
| Input and output layer dimensions | 128 × 12 and 128 × 12 |
| Number of epochs | 500 |
| *Layer* | *Hyperparameters* |
| Encoder Module: 1st Convolutional module | Conv2d (27 channels, kernel size=3, stride=2, padding=1), LeakyReLU(0.01), Dropout(0.119353206904521) |
| Flatten | Flatten (27 × 64 × 6) |
| 1st Fully connected module | Linear (10368, 89), LeakyReLU(0.01) |
| 2nd Fully connected module | Linear (89, 10368), LeakyReLU(0.01) |
| Unflatten | Unflatten (27 × 64 × 6) |
| Decoder Module: 1st Deconvolutional module | ConvTranspose2d (1 channels, kernel size=3, stride=2, padding=1, output padding=1), LeakyReLU(0.01) |
| Phase 2: Mapping ($G_{A1}$) | |
| Initial learning rate | 0.0000948439065685934 |
| Loss | Mean squared error |
| Input and output layer dimensions | 89 and 42 |
| Number of epochs | 100 |
| *Layer* | *Hyperparameters* |
| 1st Fully connected module | Linear (89, 67), ReLU(), Dropout (0.09572) |
| 2nd Fully connected module | Linear (67, 587), ReLU(), Dropout (0.09572) |
| 3rd Fully connected module | Linear (587, 42) |
| Phase 3: Finetuning ($G_{A2}$) | |
| Model | Logistic Regression Classifier |
| Regularization Penalty | L2 norm |
| Classification Loss | Binary cross entropy |
| Solver | lbfgs |
| Input and output dimensions | 42 and 1 |
| *Hyperparameters* | *Values* |
| Tolerance for stopping criteria | 0.0001 |
| Number of iterations | 150 |
| Random state | 42 |